\documentclass[journal]{IEEEtran}

\usepackage{amsthm}
\usepackage{graphicx}
\usepackage{booktabs}
\usepackage{cite}
\usepackage{bbding}
\usepackage{amsmath}
\usepackage{amsfonts}
\usepackage{cases}
\usepackage{algorithmic}
\usepackage{algorithm}
\usepackage{array}
\usepackage{multirow}
\usepackage{threeparttable}
\usepackage{bm}
\usepackage[colorlinks,linkcolor=black, citecolor=blue]{hyperref}
\usepackage[version=4]{mhchem}
\usepackage{amsmath,nccmath}

\begin{document}

\title{Accelerated MRI Reconstruction with Separable and Enhanced Low-Rank Hankel Regularization}%

\author{Xinlin Zhang, Hengfa Lu, Di Guo, Zongying Lai, Huihui Ye, Xi Peng, Bo Zhao, and Xiaobo Qu

\thanks{This work was supported in part by National Natural Science Foundation of China (61971361, 61871341, 61811530021, U1632274, and 61901188), National Key R\&D Program of China (2017YFC0108703), Health Education Joint Project of Fujian Province (2019-WJ-31), and Xiamen University Nanqiang Outstanding Talents Program. \emph{Corresponding author: Xiaobo Qu}.}%
\thanks{X. Zhang and X. Qu are with Biomedical Intelligent Cloud R\&D Center, Department of Electronic Science, National Institute for Data Science in Health and Medicine, Xiamen University, Xiamen 361105 China (e-mail: quxiaobo@xmu.edu.cn).}
\thanks{Hengfa Lu is with the Department of Biomedical Engineering, University of Texas at Austin, Austin, TX 78712 USA.}
\thanks{D. Guo is with the School of Computer and Information Engineering, Xiamen University of Technology, Xiamen 361021 China.}
\thanks{Z. Lai is with the School of Information Engineering, Jimei University, Xiamen 361024 China.}
\thanks{H. Ye is with the State of Key Laboratory of Modern Optical Instrumentation, College of Optical Science and Engineering, Zhejiang University, Hangzhou 310058 China.}
\thanks{X. Peng is with the Department of Radiology, Mayo Clinic, Rochester, MN 55902 USA.}
\thanks{B. Zhao is with the Department of Biomedical Engineering, Oden Institute for Computational Engineering and Sciences, University of Texas at Austin, Austin, TX 78712 USA.}}

\maketitle

%----------------------------------------------------------------------
%----------------------------- Abstract -------------------------------
%----------------------------------------------------------------------
\begin{abstract}
The combination of the sparse sampling and the low-rank structured matrix reconstruction has shown promising performance, enabling a significant reduction of the magnetic resonance imaging data acquisition time. However, the low-rank structured approaches demand considerable memory consumption and are time-consuming due to a noticeable number of matrix operations performed on the huge-size block Hankel-like matrix. In this work, we proposed a novel framework to utilize the low-rank property but meanwhile to achieve faster reconstructions and promising results. The framework allows us to enforce the low-rankness of Hankel matrices constructing from 1D vectors instead of 2D matrices from 1D vectors and thus avoid the construction of huge block Hankel matrix for 2D k-space matrices. Moreover, under this framework, we can easily incorporate other information, such as the smooth phase of the image and the low-rankness in the parameter dimension, to further improve the image quality. We built and validated two models for parallel and parameter magnetic resonance imaging experiments, respectively. Our retrospective \textit{in-vivo} results indicate that the proposed approaches enable faster reconstructions than the state-of-the-art approaches, e.g., about $8 \times$ faster than STDLR-SPIRiT, and faithful removal of undersampling artifacts.
\end{abstract}


\IEEEpeerreviewmaketitle

%----------------------------------------------------------------------
%--------------------------- Introduction -----------------------------
%----------------------------------------------------------------------
\section{Introduction}\label{Section:introduction}
\IEEEPARstart{M}{agnetic} resonance imaging (MRI) is a non-radioactive, non-invasive imaging technique that is able to provide multi-contrast images and permit excellent soft-tissue images. It has become an indispensable tool in medical diagnosis. However, the long acquisition time is one of its prominent limitations, which may introduce motion-caused artifacts into the images and may not be practical in some application scenarios \cite{2017_review_PI}.

To alleviate the prolonged acquisition time, researchers have made great efforts. Shortening the scan time by acquire only a small subset of the k-space data has emerged as an effective and widely-used way \cite{2007_MRM_Sparse_MRI,2007_PSF_Liang}. The sparse sampling technique enables significant reduction of time, but results in undersampling artifacts in the images due to sub-Nyquist sampling. Therefore, reconstruction models with prior information are exploited to enable promising results. Sparsity, low-rank, and sparsity plus low-rank are commonly-used priors. For instance, the sparse MRI methods enforce the sparsity of images in the transform domain, e.g., total variation \cite{2007_MRI_TV}, wavelets \cite{2007_MRM_Sparse_MRI}, contourlets \cite{2010_xiaobo}, and adaptive sparse transform \cite{2011_Ravishankar,2014_PANO,2016_TBME_Zhifang,2016_MIA_Lai,2016_Data_driven_Ravishankar,2016_pFISTA}. The low-rank prior stems from the linear correlations among multiple MRI images \cite{2007_PSF_Liang}, such as, dynamic MRI \cite{2012_TMI_Zhao,2015_LR_sparse_MRI}, parameter imaging \cite{2015_mapping,2016_MORASA}, and high-dimensional MRI \cite{2015_MIA_Zhang,2016_high_dim_MR_tensor} and has been utilized.

The low-rank structured matrix prior has been shown to be able to permit promising results \cite{2014_LORAKS,2014_MRM_Lustig,2016_ALOHA,2016_Off_the_grid,2020_xinlin}. For instance, simultaneous autocalibrating and k-space estimation \cite{2014_MRM_Lustig} utilizes the strong correlation among receiver coils to enable the low-rankness of a block Hankel matrix constructed from the multi-coil k-space data. The low-rank matrix modeling of local k-space neighborhoods (LORAKS) \cite{2014_LORAKS} assumes the image of interest has limited support and/or a slowly varying phase, which would lead to a block-Hankel-like matrix being low-rank. The annihilating filter-based low-rank Hankel matrix approach (ALOHA) \cite{2016_ALOHA}, and Ongie’s method \cite{2016_Off_the_grid} consider the sparsity of the signal in the transform domain. Similarly, the simultaneous two-directional low-rankness with SPIRiT (STDLR-SPIRiT) method \cite{2020_xinlin} exploits the low-rank prior and enforces the self-consistency of k-space data providing state-of-the-art performance. 

The structured low-rank methods lift the signal into a higher dimensional space to achieve low-rankness. However, the lifting brings some limitations, such as, large memory consumption and lengthy computational time. As for four-coil $256 \times 256$ data, the dimension of its block Hankel matrix reaches $2116 \times 54756$ when the pencil parameters equal $23$. Researcher have established methods aiming to tackle the problem. The  generalized iterative reweighted annihilating filter algorithm \cite{2017_GIRAF} utilized the convolutional structure of the lifted matrix so that it can performs the computations in the original signal instead of the lifted matrix. This method dramatically reduces the complexity and shortens the computational time. Auto-calibrated-LORAKS (AC-LORAKS) \cite{2015_AC-LORAKS} exploited the auto-calibration signal (ACS) to reduce the computational time. Though the two methods gain considerable acceleration in time, they take advantage of specific priors that may not available in other low-rank approaches. Thus, fast low-rank Hankel MRI reconstruction is still highly desirable.

In this work, we attempted to establish a framework to permit fast and reliable low-rank reconstructions. A separable low-rank reconstruction method was proposed by enforcing the low-rankness of multiple small Hankel matrices from each row and each column of MRI data, avoiding constructing the high-dimensional low-rank matrix thereby enables much less memory and allows faster computation. In addition, the self-consistence and conjugate symmetric of k-space data are considered to further reduce the reconstruction error. Besides, the parameter-dimensional information will be introduced into formulation so that the original model can not only be applied to non-parameter parallel imaging but also extended to parameter imaging. The proposed method is expected to effectively utilize the low-rank property to yield fast reconstruction and maintain low reconstruction error.

 The rest of this paper is organized as follows. We introduce some notations in Section \ref{Section:notations}, and related work in Section \ref{Section:relatedWork}. Section \ref{Section:method} presents the proposed model and numerical algorithm for non-parameter and parameter parallel imaging reconstructions, and Section III demonstrates the reconstruction performance. Section \ref{Section:discussion} discusses some factors that may affect the reconstruction error and speed. The conclusions are finally drawn in Section \ref{Section:conclusion}.

\begin{figure}[htbp]
\setlength{\abovecaptionskip}{0.cm}
\setlength{\belowcaptionskip}{-0.cm}
\centering
\includegraphics[width=3.4in]{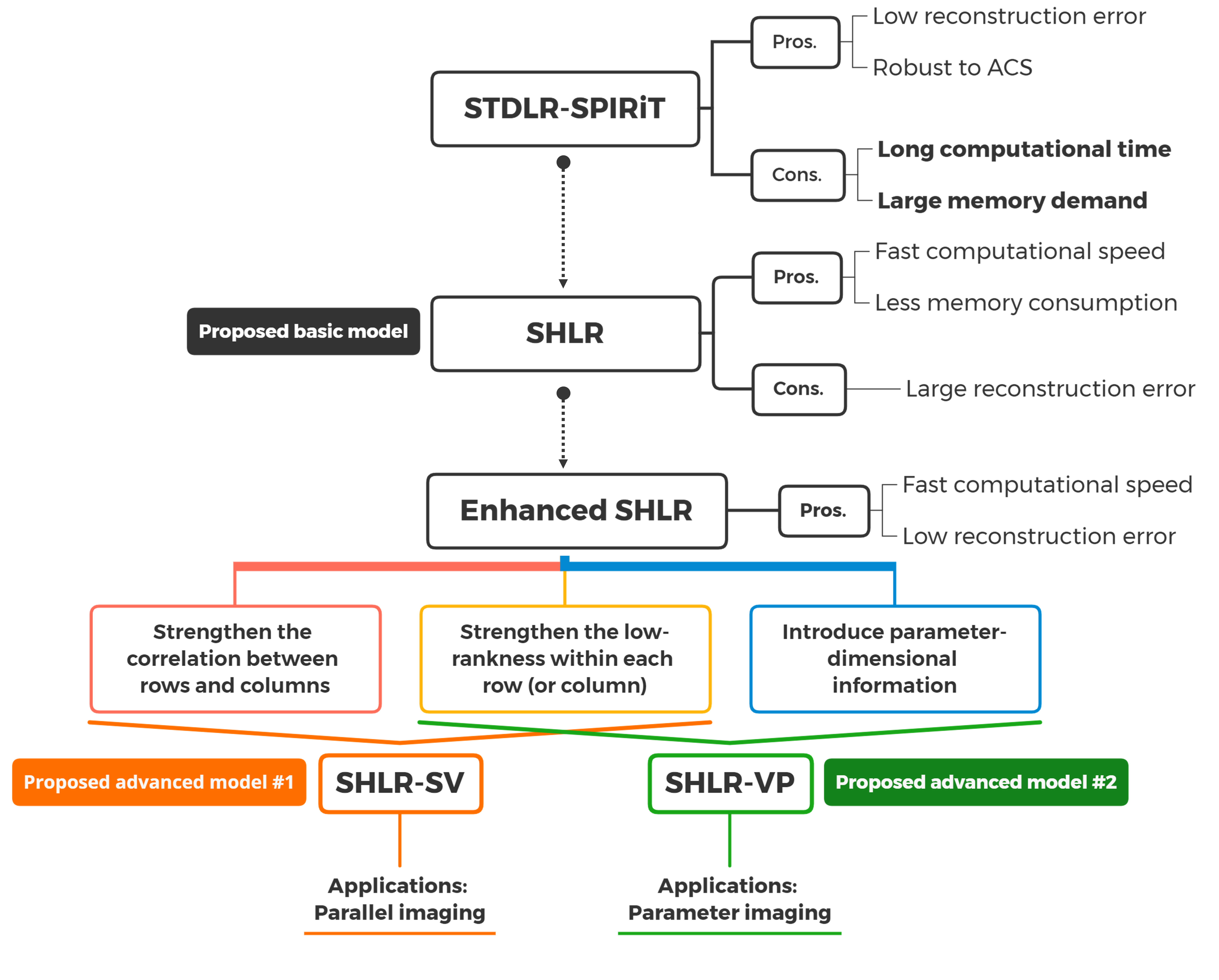}
\caption{Mind map of this work.}
\label{fig_mindmap}
\end{figure}

%----------------------------------------------------------------------
%----------------------------- Notations ------------------------------
%----------------------------------------------------------------------
\section{Notations} \label{Section:notations}
For easier reading, we first list some notations used throughout this paper in Table \ref{Table_notation}.

\begin{table}[htbp]
  \centering
  \caption{Notations used in this paper.}
    \begin{tabular}{m{1.5cm}m{6.5cm}}
    \toprule
    Symbol & Quantity \\
    \midrule
    $\mathbf{K}$     &  desired multi-coil k-space data, $M$ (rows) $\times$ $N$ (columns) $\times$ $J$ (coils)\\
    $\mathbf{X}$     &  desired multi-coil image, $M$ (rows) $\times$ $N$ (columns) $\times$ $J$ (coils)\\
    $\mathbf{x}^{\text{row}}_{m,j}$     & vector of the $m$-th row and the $j$-th coil of $\mathbf{X}$\\
    $\mathbf{x}^{\text{col}}_{n,j}$     & vector of the $n$-th column and the $j$-th coil of $\mathbf{X}$ \\
    $\mathbf{X}^{\text{Parameter}}$     & desired image of parameter imaging , $M$ (FE) $\times$ $N$ (PE) $\times$ $L$ (P) $\times$ $J$ (coils)\\
    $\mathbf{X}^{\text{PE-P}}_{m}$     & desired image at the $m$-th position on FE dimension of parameter imaging, $N$ (PE) $\times$ $L$ (P) $\times$ $J$ (coils)\\
    \bottomrule
    \end{tabular}%
    \begin{tablenotes}[flushleft]
    \footnotesize
    \item Note: the FE is the abbreviation of frequency encoding, the PE is the abbreviation of phase encoding, the capital letter P denotes the parameter dimension.
     \end{tablenotes}
  \label{Table_notation}%
\end{table}

%----------------------------------------------------------------------
%---------------------------- Related Work ----------------------------
%----------------------------------------------------------------------
\section{Related Work} \label{Section:relatedWork}
The STDLR \cite{2020_xinlin} simultaneously enforced the low-rank constraint of k-space data along both horizontal and vertical directions as follows:
\begin{equation}
\underset{\mathbf{K}}{\mathop{\min }}{{\left\| \mathbf{BW}_{\bot }^{2\text{D}}\mathbf{K} \right\|}_{*}} \! + \! {{\left\| \mathbf{BW}_{=}^{2\text{D}}\mathbf{K} \right\|}_{*}} \! + \! \frac{\lambda }{2} \! \left\| \text{vec}\left( \mathbf{Y} \!-\! \mathbf{UK} \right) \right\|_{F}^{2},
\end{equation}
where $\mathbf{K}\in {{\mathbb{C}}^{M\times N\times J}}$ denotes the targeted k-space data, and $\mathbf{Y}\in {{\mathbb{C}}^{M\times N\times J}}$ the acquired multi-coil k-space data with zero-filling in unacquired position. The operator $\mathbf{B}$ transfers multi-coil data into a cascaded block Hankel matrix; $\mathbf{W}_{=}^{2\text{D}}$ and $\mathbf{W}_{\bot }^{2\text{D}}$ are weighting operators that perform weighting on k-space data of each coil image, whose weights are the Fourier transform of filters in the horizontal and vertical directions; $\mathbf{U}$ represents an operator that undersamples data, and $\text{vec}\left( \cdot  \right)$ means arranging a matrix or a tensor to a vector.

\begin{figure}[htbp]
\setlength{\abovecaptionskip}{0.cm}
\setlength{\belowcaptionskip}{-0.cm}
\centering
\includegraphics[width=3.4in]{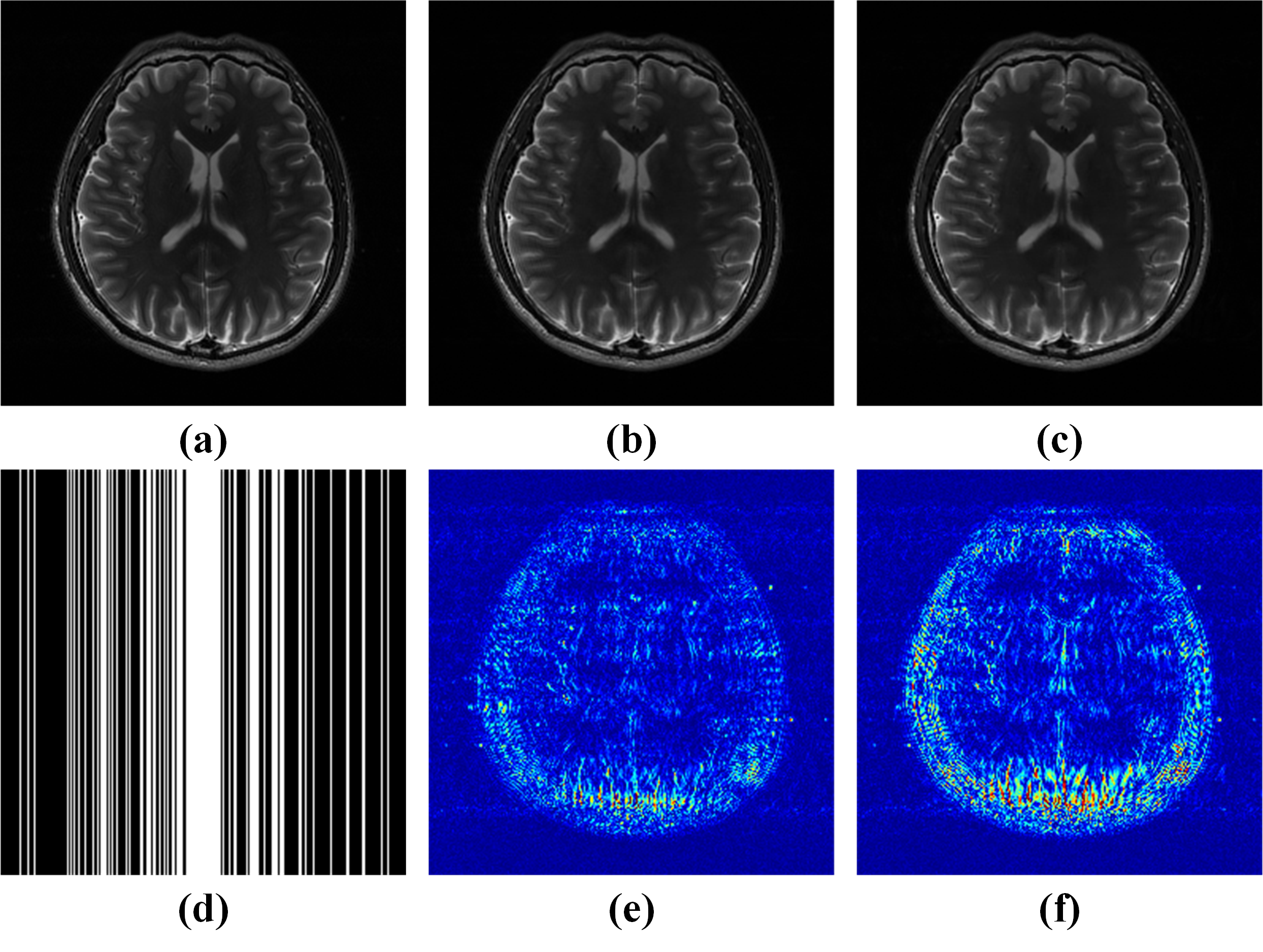}
\caption{Reconstructions of a brain image using STDLR and SHLR. (a) The fully sampled SSOS image; (b-c) SSOS images of reconstructed results by STDLR and SHLR, respectively; (d) the Cartesian undersampling pattern with a sampling rate of 0.34; (e-f) the reconstruction error distribution (12.5$\times$) corresponding to the above methods. Note: the RLNE of (b-c) are 0.0611 and 0.0825, and the computational time of (b-c) are 1014.2 s and 40.8 s, respectively.}
\label{fig_method_2}
\end{figure}

The STDLR explored the simultaneous horizontal and vertical directional low-rankness in the k-space data, reducing the image reconstruction errors. Compared to the state-of-the-art method ALOHA, STDLR achieves lower reconstruction errors \cite{2020_xinlin}. However, the size of the block Hankel matrix appears dramatically huge, leading to a long computational time and large memory requirements. Take four-coil parallel imaging data for an example, the STDLR requires $1014.2$ s to finish reconstruction while the separable Hankel low-rank method takes only $40.8$ s (Fig. \ref{fig_method_2}).

%----------------------------------------------------------------------
%--------------------------- Proposed Method --------------------------
%----------------------------------------------------------------------

\begin{figure*}[htbp]
\setlength{\abovecaptionskip}{0.cm}
\setlength{\belowcaptionskip}{-0.cm}
\centering
\includegraphics[width=6.8in]{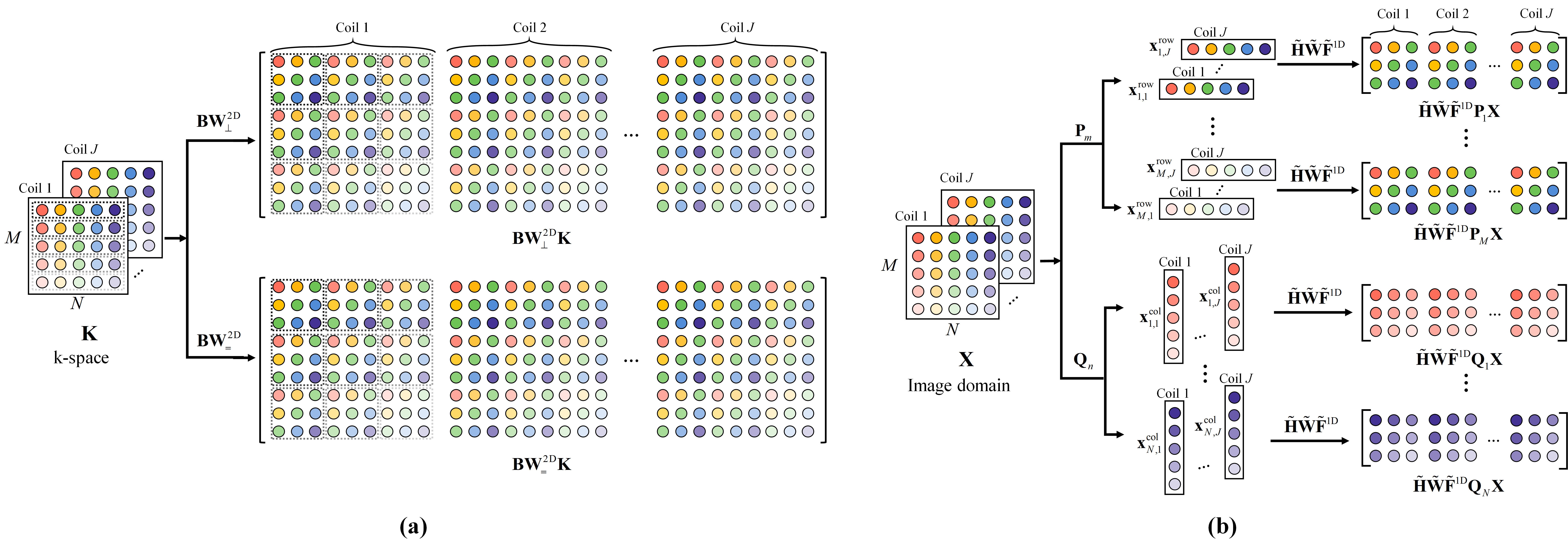}
\caption{Illustration of constructing Hankel matrices in (a) STDLR and (b) SHLR.}
\label{fig_flowchart}
\end{figure*}

\section{Proposed Method} \label{Section:method}
\subsection{Basic Model - SHLR}
In this work, we first proposed a separable Hankel low-rank reconstruction method (SHLR) to enforce the low-rankness of Hankel matrices constructing from each row and each column as follows:
\begin{equation}
\begin{medsize}
\begin{aligned}
\textbf{(SHLR)} \quad \underset{\mathbf{X}}{\mathop{\min }} & \sum\limits_{m=1}^{M} {{{\left\| \mathbf{\tilde{H}\tilde{W}}{{{\mathbf{\tilde{F}}}}^{\text{1D}}}{{\mathbf{P}}_{m}}\mathbf{X} \right\|}_{*}}} + \sum\limits_{n=1}^{N}{{{\left\| \mathbf{\tilde{H}\tilde{W}}{{{\mathbf{\tilde{F}}}}^{\text{1D}}}{{\mathbf{Q}}_{n}}\mathbf{X} \right\|}_{*}}} \\
& \qquad \qquad \qquad \qquad \quad + \frac{\lambda }{2}\left\| \text{vec}\left( \mathbf{Y}-\mathbf{U}{{\mathbf{F}}^{\text{2D}}}\mathbf{X} \right) \right\|_{2}^{2},
\end{aligned}
\end{medsize}
\end{equation}
 where the $\mathbf{X}\in {{\mathbb{C}}^{M\times N\times J}}$ denotes the desired multi-coil image, ${{\mathbf{F}}^{\text{2D}}}$ is the 2D Fourier transform for each coil data. The ${{\mathbf{P}}_{m}}$ and ${{\mathbf{Q}}_{n}}$ denote the operators that extract $m$-th row ($m$-th vector on the first dimension) and $n$-th column ($n$-th vector on the second dimension) from each coil data for $m=1,\cdots ,M$ and $n=1,\cdots ,N$. Here, we define $\mathbf{x}_{m,j}^{\text{row}}$ as the $m$-th row and $j$-th coil of $\mathbf{X}$, and $\mathbf{x}_{n,j}^{\text{col}}$ as the $n$-th column and $j$-th coil of $\mathbf{X}$. Then, we have ${{\mathbf{P}}_{m}}\mathbf{X}=\left[ \mathbf{x}_{m,1}^{\text{row}},\cdots ,\mathbf{x}_{m,j}^{\text{row}},\cdots ,\mathbf{x}_{m,J}^{\text{row}} \right]\in {{\mathbb{C}}^{N\times J}}$ and ${{\mathbf{Q}}_{n}}\mathbf{X}=\left[ \mathbf{x}_{n,1}^{\text{col}},\cdots ,\mathbf{x}_{n,j}^{\text{col}},\cdots ,\mathbf{x}_{n,J}^{\text{col}} \right]\in {{\mathbb{C}}^{M\times J}}$.

 Fig. \ref{fig_flowchart} shows a graphical illustration of the notations of constructing Hankel matrices. The operator ${{\mathbf{F}}^{1\text{D}}}$ denotes the 1D Fourier transform on a vector, $\mathbf{W}$ performs weighting on a vector with the weights obtained from applying Fourier transform to 1D sparse transform filter, and $\mathbf{H}$ converts a vector into a Hankel matrix. The tilde above the operator means that corresponding operation is performed on each column vector of the matrix, that is $\mathbf{\tilde{H}\tilde{W}}{{\mathbf{\tilde{F}}}^{\text{1D}}}{{\mathbf{P}}_{m}}\mathbf{X}=\left[ \mathbf{HW}{{\mathbf{F}}^{\text{1D}}}\mathbf{x}_{m,1}^{\text{row}},\cdots ,\mathbf{HW}{{\mathbf{F}}^{\text{1D}}}\mathbf{x}_{m,J}^{\text{row}} \right]$ and $\mathbf{\tilde{H}\tilde{W}}{{\mathbf{\tilde{F}}}^{1\text{D}}}{{\mathbf{Q}}_{n}}\mathbf{X}=\left[ \mathbf{HW}{{\mathbf{F}}^{1\text{D}}}\mathbf{x}_{n,1}^{\text{col}},\cdots ,\mathbf{HW}{{\mathbf{F}}^{1\text{D}}}\mathbf{x}_{n,J}^{\text{col}} \right]$.

\subsection{Enhanced Model} 
As mentioned above, building the block Hankel matrix requires huge memory consumption and thereby increases the computational complexity and time. To mitigate the problem, we proposed to establish multiple small-size Hankel matrices by exploiting the low-rankness of a row or a column vector of the signal. This will significantly reduce memory consumption and make computational speed much faster.

However, constraining the low-rankness of each row and each column separately weakens the utilization between rows and columns, leading to an increase in reconstruction errors, e.g. relative $\ell_2$ norm error (RLNE). (Fig. \ref{fig_method_2} (e-f)). In order to improve the reconstruction while maintaining a faster reconstruction speed, we consider introducing other information to strengthen the constraints of the model. The following three kinds of information are exploited.

\subsubsection{Strengthen the correlation between rows and columns}
The proposed SHLR method makes little use of the correlation between rows and columns. A natural idea for improvement is to strengthen the exploitation of k-space neighborhood information. The SPIRiT is a good choice. The SPIRiT \cite{2010_SPIRiT} primarily bases on the assumption that each k-space data point is the convolution of multi-coil data of its neighboring k-space, which implies the SPIRiT constraint has the ability to utilize the linear correlation of adjacent rows and columns to reconstruct MRI image. Therefore, the SHLR with SPIRiT constraint (SHLR-S) can be modeled as:
\begin{equation}
\begin{medsize}
\begin{aligned}
& \textbf{(SHLR-S)} \;\; \underset{\mathbf{X}}{\mathop{\min }} \sum\limits_{m=1}^{M}{{{\left\| \mathbf{\tilde{H}\tilde{W}}{{{\mathbf{\tilde{F}}}}^{\text{1D}}}{{\mathbf{P}}_{m}}\mathbf{X} \right\|}_{*}}}+\sum\limits_{n=1}^{N}{{{\left\| \mathbf{\tilde{H}\tilde{W}}{{{\mathbf{\tilde{F}}}}^{\text{1D}}}{{\mathbf{Q}}_{n}}\mathbf{X} \right\|}_{*}}}\\
&\qquad \qquad \ +\frac{\lambda }{2}\left\| \text{vec}\left( \mathbf{Y}-\mathbf{U}{{\mathbf{F}}^{\text{2D}}}\mathbf{X} \right) \right\|_{2}^{2}+\frac{{{\lambda }_{1}}}{2}\left\| \text{vec}\left( \mathbf{X}-\mathbf{GX} \right) \right\|_{2}^{2},
\end{aligned}
\end{medsize}
\end{equation}
where $\mathbf{G}$ denotes the SPIRiT operator in image domain.

\subsubsection{Strengthen the low-rankness within each row (or column)}
Another way to improve the reconstruction is to strengthen the low-rankness within each row and each column. The smooth phase of MRI images is often used as prior information to be involved in MRI reconstruction \cite{1991_PF,2012_SmoothPhase,2014_LORAKS}. By generating the virtual coils containing the conjugate symmetric, the smooth MRI image phase information can be incorporated into the reconstruction problem \cite{2009_VC}. The conjugate symmetry of k-space has been applied into the Hankel low-rank reconstruction \cite{2020_Hankel_VC}. This property can be easily introduced into SHLR reconstruction model as:
\begin{equation}
\begin{medsize}
\begin{aligned}
   {{{\mathbf{\tilde{H}}}}_{\text{vc}}}\mathbf{\tilde{W}}{{{\mathbf{\tilde{F}}}}^{\text{1D}}}{{\mathbf{P}}_{m}}\mathbf{X} & =\left[ \mathbf{HW}{{\mathbf{F}}^{\text{1D}}}\mathbf{x}_{m,1}^{\text{row}},\cdots ,\mathbf{HW}{{\mathbf{F}}^{\text{1D}}}\mathbf{x}_{m,J}^{\text{row}}, \right.\\
 & \left. \mathbf{HW}{{\left( {{\mathbf{F}}^{\text{1D}}}\mathbf{x}_{m,1}^{\text{row}} \right)}^{\dagger }},\cdots ,\mathbf{HW}{{\left( {{\mathbf{F}}^{\text{1D}}}\mathbf{x}_{m,J}^{\text{row}} \right)}^{\dagger }} \right], \\ 
  {{{\mathbf{\tilde{H}}}}_{\text{vc}}}\mathbf{\tilde{W}}{{{\mathbf{\tilde{F}}}}^{\text{1D}}}{{\mathbf{Q}}_{n}}\mathbf{X} & =\left[ \mathbf{HW}{{\mathbf{F}}^{\text{1D}}}\mathbf{x}_{n,1}^{\text{col}},\cdots ,\mathbf{HW}{{\mathbf{F}}^{\text{1D}}}\mathbf{x}_{n,J}^{\text{col}}, \right.\\
 & \left. \mathbf{HW}{{\left( {{\mathbf{F}}^{\text{1D}}}\mathbf{x}_{n,1}^{\text{col}} \right)}^{\dagger }},\cdots ,\mathbf{HW}{{\left( {{\mathbf{F}}^{\text{1D}}}\mathbf{x}_{n,J}^{\text{col}} \right)}^{\dagger }} \right]. 
\end{aligned}
\end{medsize}
\end{equation}
where the superscript $\dagger$ represents the operation of taking the conjugate and flipping the vector along the center. Therefore, we can formulate the SHLR with virtual coil (SHLR-V) as:
{\small
\begin{multline}
\textbf{(SHLR-V)} \quad \quad \underset{\mathbf{X}}{\mathop{\min }} \frac{\lambda }{2}\left\| \text{vec}\left( \mathbf{Y}-\mathbf{U}{{\mathbf{F}}^{\text{2D}}}\mathbf{X} \right) \right\|_{2}^{2} + \\
\sum\limits_{m=1}^{M}{{{\left\| {{{\mathbf{\tilde{H}}}}_{\text{vc}}}\mathbf{\tilde{W}}{{{\mathbf{\tilde{F}}}}^{\text{1D}}}{{\mathbf{P}}_{m}}\mathbf{X} \right\|}_{*}}}+\sum\limits_{n=1}^{N}{{{\left\| {{{\mathbf{\tilde{H}}}}_{\text{vc}}}\mathbf{\tilde{W}}{{{\mathbf{\tilde{F}}}}^{\text{1D}}}{{\mathbf{Q}}_{n}}\mathbf{X} \right\|}_{*}}}.
\end{multline} }

We introduced both the SPIRiT and virtual coil information into the basis SHLR model aiming to further reduce the reconstruction errors. We can achieve better reconstruction results with the optimization as:
\begin{equation}\label{(SHLR-SV_model)}
\begin{medsize}
\begin{aligned}
 \textbf{(SHLR-SV)} \; \underset{\mathbf{X}}{\mathop{\min }}\frac{\lambda }{2}\left\| \text{vec} \! \left( \! \mathbf{Y} \! - \! \mathbf{U}{{\mathbf{F}}^{\text{2D}}}\mathbf{X} \! \right) \! \right\|_{2}^{2} \! + \! \frac{{{\lambda }_{1}}}{2} \! \left\| \text{vec}\left( \! \mathbf{X} \! - \! \mathbf{GX} \right) \right\|_{2}^{2}\\
+\sum\limits_{m=1}^{M}{{{\left\| {{{\mathbf{\tilde{H}}}}_{\text{vc}}}\mathbf{\tilde{W}}{{{\mathbf{\tilde{F}}}}^{\text{1D}}}{{\mathbf{P}}_{m}}\mathbf{X} \right\|}_{*}}} \! + \! \sum\limits_{n=1}^{N}{{{\left\| {{{\mathbf{\tilde{H}}}}_{\text{vc}}}\mathbf{\tilde{W}}{{{\mathbf{\tilde{F}}}}^{\text{1D}}}{{\mathbf{Q}}_{n}}\mathbf{X} \right\|}_{*}}}.
\end{aligned}
\end{medsize}
\end{equation}

\begin{figure}[htbp]
\setlength{\abovecaptionskip}{0.cm}
\setlength{\belowcaptionskip}{-0.cm}
\centering
\includegraphics[width=3.4in]{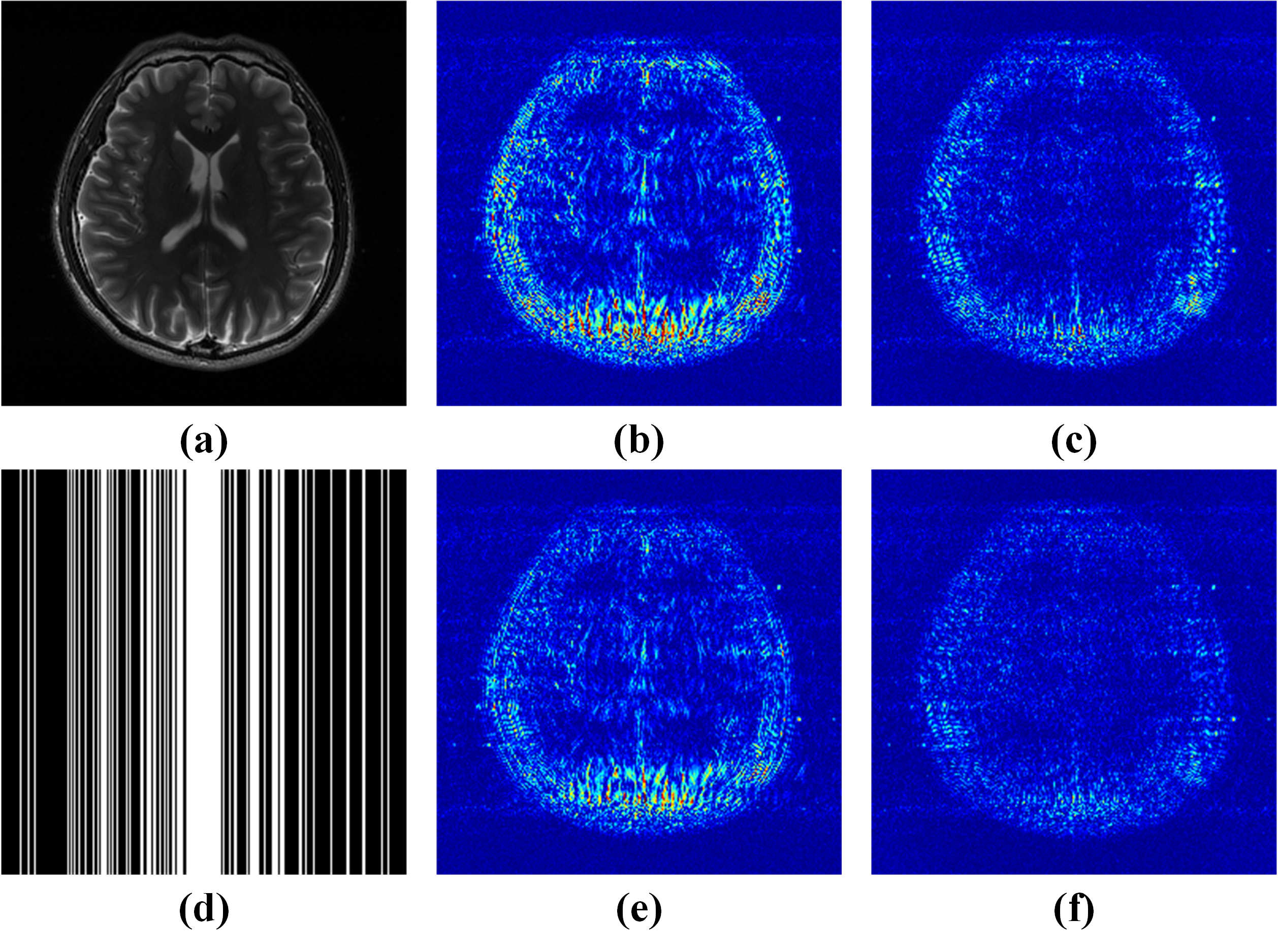}
\caption{Reconstruction on brain image using SHLR and its variants. (a) The fully sampled SSOS image; (b, c, e, f) the reconstruction error (12.5$\times$) of SHLR, SHLR-S, SHLR-V, and SHLR-SV, respectively; (d) the Cartesian undersampling pattern with a sampling rate of 0.34. Note: the RLNE of (b, c, e, f) are 0.0825, 0.0527, 0.0664, and 0.0426; and the corresponding computational time are 40.8 s, 84.6 s, 60.2 s, and 97.6 s.}
\label{fig_method_1}
\end{figure}

As shown in Fig. \ref{fig_method_1}, the SHLR-SV, which equipped with both the SPIRiT and virtual coil information, produces the image with lowest error (Fig. \ref{fig_method_1} (f)) while the basic SHLR and the SHLR with only one extra information enforcement yield results with obvious errors (Fig. \ref{fig_method_1} (b-c) and (e)). The results indicate the effectiveness of utilizing the two information in parallel imaging.

\subsubsection{Parameter-dimensional information}
Extra information can be incorporated to improve reconstructions of multi-dimensional MRI, e.g., information from parameter imaging.

Conventional parameter imaging acquires a series of images using different parameters. The image intensity variation along the parameter dimension then is used in data fitting to estimate the tissue intrinsic parameters, such as longitudinal relaxation time (T1), transversal relaxation time (T2). The tissue parameters are important in diagnosing the diseases including nervous, musculoskeletal, liver, and myocardial \cite{2008_mapping,2010_mapping,2011_mapping,2011_Feng_mapping,2014_Bo_parameter,2015_mapping}. It is worthy to mention that the signal along the parameter dimension can be modeled as mono or few exponentials \cite{2010_mapping,2014_Bo_parameter,2016_ALOHA_mapping}, which indicates the low-rankness of the signal \cite{2016_MORASA,2016_ALOHA_mapping,2017_Shah}. The low-rank property along the parameter dimension can be utilized to enhance the reconstructions.

\begin{figure}[!htb]
\setlength{\abovecaptionskip}{0.cm}
\setlength{\belowcaptionskip}{-0.cm}
\centering
\includegraphics[width=3.4in]{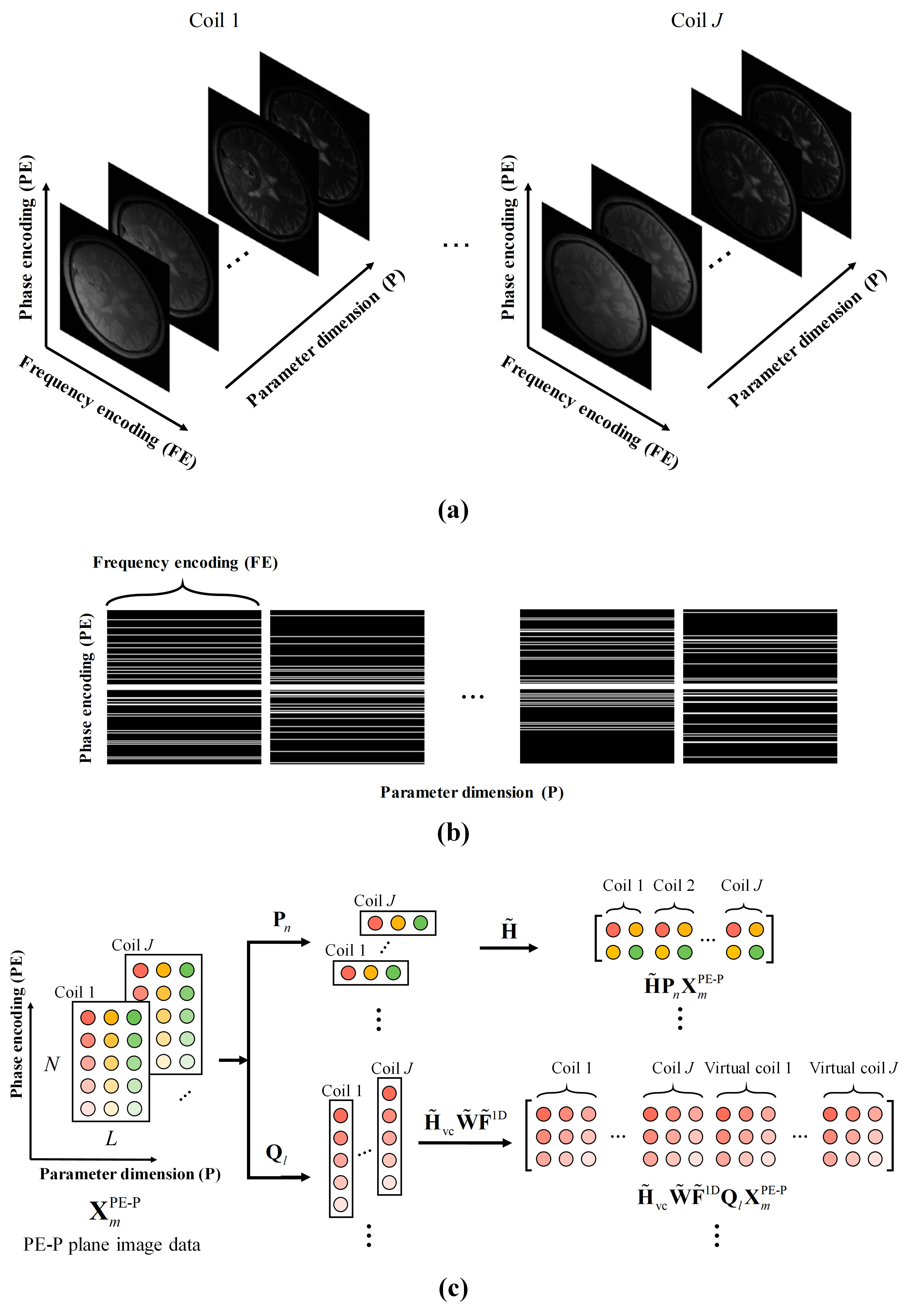}
\caption{Graphical illustration of proposed parameter imaging reconstruction. (a) the diagrammatic sketch of parameter imaging data; (b) the undersampling patterns; (c) constructing the Hankel matrices of proposed method SHLR-VP.}
\label{fig_flowchart_mapping}
\end{figure}

\begin{figure}[!htb]
\setlength{\abovecaptionskip}{0.cm}
\setlength{\belowcaptionskip}{-0.cm}
\centering
\includegraphics[width=2.8in]{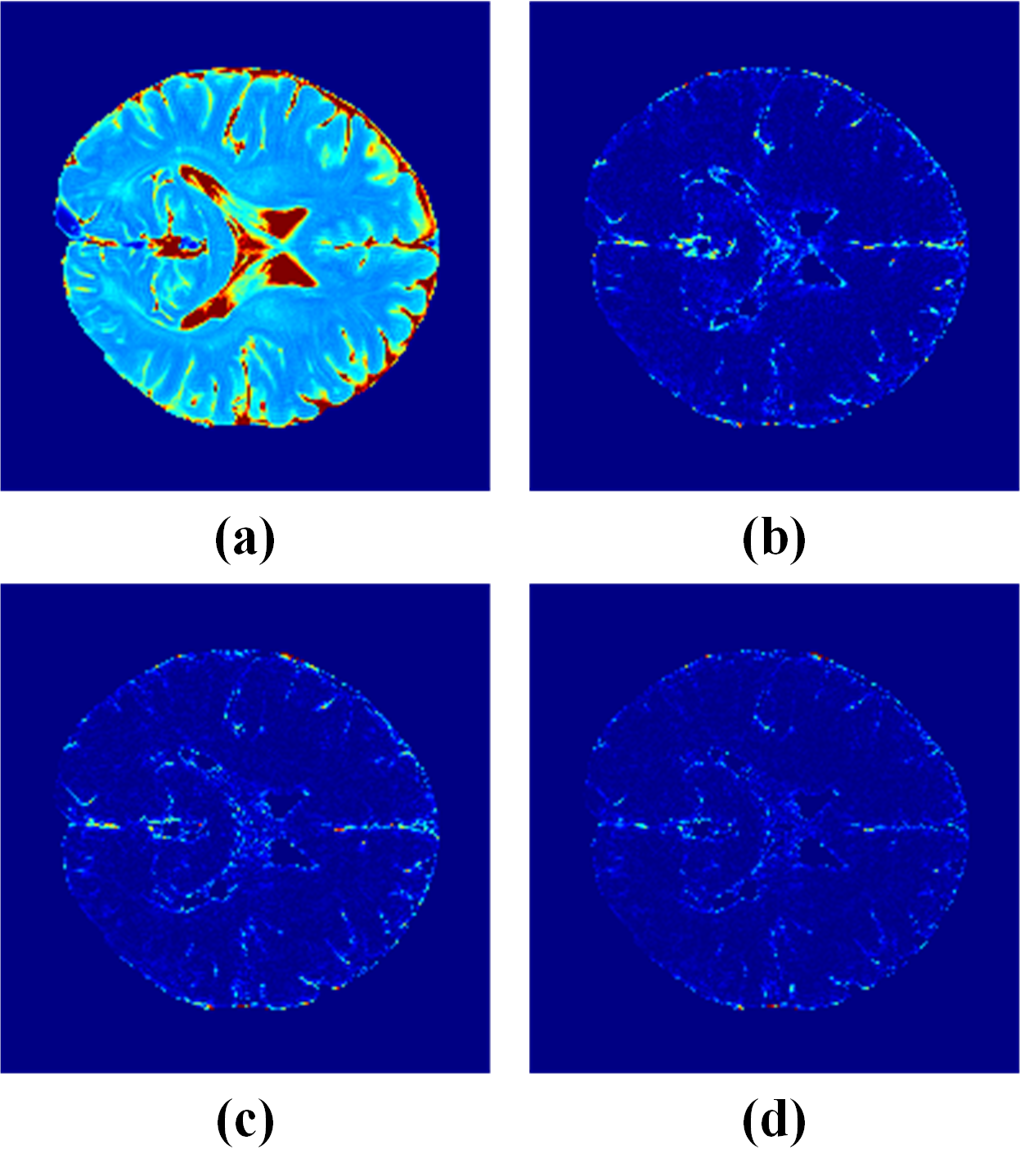}
\caption{Reconstruction on T2 mapping using ALOHA and the proposed method with the reduction factor R=8. (a) The T2 map estimated from fully sampled data; (b-d) the T2 map error distribution (6$\times$) of ALOHA, SHLR-P, and SHLR-VP, respectively. Note: the RLNE of (b-d) are 0.1262, 0.1174, and 0.1029; and the computational time of (b-d) are 309 s, 108 s, and 197 s, respectively.}
\label{fig_SHLR-P}
\end{figure}

Motivated by ALOHA \cite{2016_ALOHA_mapping}, we reconstruct images in the frequency encoding dimension one-by-one (Fig. \ref{fig_flowchart_mapping}). The difference between ALOHA and our proposed method is that ALOHA constructed a block Hankel matrix from the data on the phase encoding-parameter (PE-P) plane while the proposed method converted rows or columns on the phase encoding-parameter plane into Hankel matrices separately. The seperable low-rank Hankel matrix strategy reduces the reconstruction time to a certain extent (Fig. \ref{fig_SHLR-P}).

Let ${{\mathbf{X}}^\text{Parameter}}\in {{\mathbb{C}}^{M\times N\times L\times J}}$ denote the desired parameter imaging MRI image. Denote $\mathbf{X}_{m}^{\text{PE-P}}\in {{\mathbb{C}}^{N\times L\times J}}$ as the image at the $m$-th position on the frequency encoding dimension of $\mathbf{X}^\text{Parameter}$. Let $\mathbf{Y}^\text{Parameter}$ denote the acquired k-space data of parameter imaging, with zero-filling at unacquired position and 1D inverse Fourier transform on along the frequency encoding dimension. And let $\mathbf{Y}_{m}^{\text{PE-P}}$ represent the data at the $m$-th position on the frequency encoding dimension of $\mathbf{Y}^\text{Parameter}$. ${{\mathbf{F}}^{\text{PE}}}$ denotes the 1D Fourier transform along the phase encoding dimension.
The reconstruction model for parameter imaging is as follow
 \begin{equation}
 \begin{medsize}
 \begin{aligned}
 \textbf{(SHLR-P)}& \quad \underset{\mathbf{X}_{m}^{\text{PE-P}}}{\mathop{\min }}\frac{\lambda }{2}\left\| \text{vec}\left( \mathbf{Y}_{m}^{\text{PE-P}}-\mathbf{U}{{\mathbf{F}}^{\text{PE}}}\mathbf{X}_{m}^{\text{PE-P}} \right) \right\|_{2}^{2} + \\
& \sum\limits_{l=1}^{L}{{{\left\| {{{\mathbf{\tilde{H}}}}}\mathbf{\tilde{W}}{{{\mathbf{\tilde{F}}}}^{1\text{D}}}{{\mathbf{Q}}_{l}}\mathbf{X}_{m}^{\text{PE-P}} \right\|}_{*}}}+{{\lambda }_{2}}\sum\limits_{n=1}^{N}{{{\left\| \mathbf{\tilde{H}}{{\mathbf{P}}_{n}}\mathbf{X}_{m}^{\text{PE-P}} \right\|}_{*}}}.
 \end{aligned}
 \end{medsize}
 \end{equation}

To further reduce the reconstruction error, we establish a reconstruction approach that takes advantages of both virtual coil and parameter-dimension information. Notice that SPIRiT exploits the correlation of image of different coils, it appears not appropriate for the recovery of PE-P plane images. Thus, the proposed method stems from the basic model and introduces two prior information, smooth phase of the image and low-rankness in the parameter dimension:
\begin{equation}\label{(SHLR-P_model)}
\begin{medsize}
\begin{aligned}
\textbf{(SHLR}&\textbf{-VP)} \quad \underset{\mathbf{X}_{m}^{\text{PE-P}}}{\mathop{\min }}\frac{\lambda }{2}\left\| \text{vec}\left( \mathbf{Y}_{m}^{\text{PE-P}}-\mathbf{U}{{\mathbf{F}}^{\text{PE}}}\mathbf{X}_{m}^{\text{PE-P}} \right) \right\|_{2}^{2} + \\
& \sum\limits_{l=1}^{L}{{{\left\| {{{\mathbf{\tilde{H}}}}_{\text{vc}}}\mathbf{\tilde{W}}{{{\mathbf{\tilde{F}}}}^{1\text{D}}}{{\mathbf{Q}}_{l}}\mathbf{X}_{m}^{\text{PE-P}} \right\|}_{*}}}+{{\lambda }_{2}}\sum\limits_{n=1}^{N}{{{\left\| \mathbf{\tilde{H}}{{\mathbf{P}}_{n}}\mathbf{X}_{m}^{\text{PE-P}} \right\|}_{*}}}.
\end{aligned}
\end{medsize}
\end{equation}

As shown in Fig. \ref{fig_SHLR-P}, the SHLR-VP which utlizes the virtual coil acheives the lowest error, and the computational time of the proposed SHLR-VP is reduced to $2/3$ of that of ALOHA. The results demonstrate the advantages of the proposed method in terms of low reconstruction error and fast reconstruction speed for parameter imaging.

\begin{figure*}[!htb]
\setlength{\abovecaptionskip}{0.cm}
\setlength{\belowcaptionskip}{-0.cm}
\centering
\includegraphics[width=6.4in]{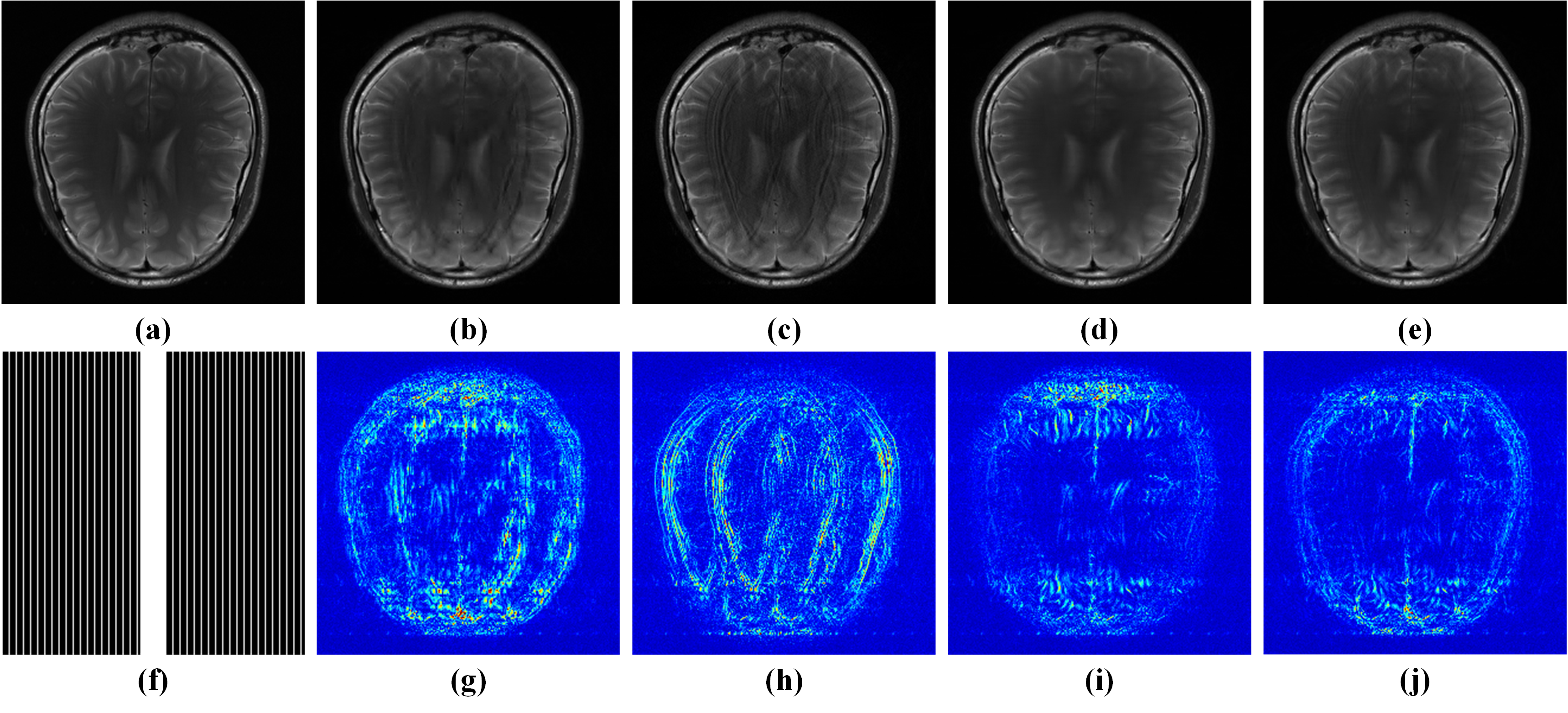}
\caption{Parallel imaging reconstruction results under uniform undersampling pattern. (a) An SSOS image of fully sampled data; (b-e) SSOS images of reconstructed results by $\ell_1$-SPIRiT, AC-LORAKS, STDLR-SPIRiT and SHLR-SV, respectively; (f) the uniform undersampling pattern with reduction factor R=6 and 20 ACS lines; (g-j) the reconstruction error distribution (10$\times$) corresponding to reconstructed image above them.}
\label{fig_2D_uniform}
\end{figure*}

\subsection{Numerical Algorithm}
We adopted the alternating direction method of multiplier (ADMM) \cite{2011_ADMM} to solve both the proposed SHLR-SV in Eq. \eqref{(SHLR-SV_model)} and the SHLR-VP model in Eq. \eqref{(SHLR-P_model)}.

\subsubsection{SHLR-SV}
The augmented Lagrangian form of Eq. \eqref{(SHLR-SV_model)} lies in
\begin{equation} \label{(SHLR-SV_ALF)}
\begin{medsize}
\begin{aligned}
  & \underset{\mathbf{D}_{m}^{\text{row}},\mathbf{D}_{n}^{\text{col}}}{\mathop{\max }}\underset{\mathbf{X},\mathbf{Z}_{m}^{\text{row}},\mathbf{Z}_{n}^{\text{col}}}{\mathop{\min }} \frac{\lambda }{2} \! \left\| \text{vec} \! \left( \! \mathbf{Y} \! -\! \mathbf{U}{{\mathbf{F}}^{\text{2D}}}\mathbf{X} \! \right) \right\|_{2}^{2} \! +\! \frac{{{\lambda }_{1}}}{2} \! \left\| \text{vec}\left( \mathbf{X} \! -\! \mathbf{GX} \right) \right\|_{2}^{2} \\ 
  & \qquad \! + \! \sum\limits_{m=1}^{M}{\left( {{\left\| \mathbf{Z}_{m}^{\text{row}} \right\|}_{*}} \! + \! \frac{\beta }{2}\left\| {{{\mathbf{\tilde{H}}}}_{\text{vc}}}\mathbf{\tilde{W}}{{{\mathbf{\tilde{F}}}}^{\text{1D}}}{{\mathbf{P}}_{m}}\mathbf{X} \! - \! \mathbf{Z}_{m}^{\text{row}} \! + \! \frac{\mathbf{D}{{_{m}^{\text{row}}}}}{\beta } \right\|_{F}^{2} \right)} \\ 
 & \qquad \! + \! \sum\limits_{n=1}^{N}{\left( {{\left\| \mathbf{Z}_{n}^{\text{col}} \right\|}_{*}} \! + \! \frac{\beta }{2}\left\| {{{\mathbf{\tilde{H}}}}_{\text{vc}}}\mathbf{\tilde{W}}{{{\mathbf{\tilde{F}}}}^{\text{1D}}}{{\mathbf{Q}}_{n}}\mathbf{X}-\mathbf{Z}_{n}^{col} \! + \! \frac{\mathbf{D}{{_{n}^{\text{col}}}}}{\beta } \right\|_{F}^{2} \right)} \\ 
\end{aligned}
\end{medsize}
\end{equation}
where $\left\langle \cdot ,\cdot  \right\rangle $ represents the inner product of matrix, and $\beta $ the penalty parameter.

The solution of Eq. \eqref{(SHLR-SV_ALF)} can be obtained by alternatively solving the following sub-problems:
\begin{equation} \label{(SHLR-SV_X)}
\begin{medsize}
\begin{aligned}
 &{{\mathbf{X}}^{\left( k+1 \right)}} \! = \! \underset{\mathbf{X}}{\mathop{\min }} \!
 \sum\limits_{m=1}^{M} \! {\frac{\beta }{2} \! \left\| {{{\mathbf{\tilde{H}}}}_{\text{vc}}}\mathbf{\tilde{W}}{{{\mathbf{\tilde{F}}}}^{\text{1D}}}{{\mathbf{P}}_{m}}{{\mathbf{X}}^{\left( k \right)}} \!\! - \! \mathbf{Z}{{_{m}^{\text{row}}}^{\left( k \right)}} \!\! +\!\! \frac{\mathbf{D}{{_{m}^{\text{row}}}^{\left( k \right)}}}{\beta } \right\|_{F}^{2}} \\ 
& \! + \! \sum\limits_{n=1}^{N}{\frac{\beta }{2}\left\| {{{\mathbf{\tilde{H}}}}_{\text{vc}}}\mathbf{\tilde{W}}{{{\mathbf{\tilde{F}}}}^{\text{1D}}}{{\mathbf{Q}}_{n}}{{\mathbf{X}}^{\left( k \right)}} \! - \! \mathbf{Z}{{_{n}^{col}}^{\left( k \right)}} \! + \! \frac{\mathbf{D}{{_{n}^{\text{col}}}^{\left( k \right)}}}{\beta } \right\|_{F}^{2}}\\
 & \! + \! \frac{\lambda }{2} \! \left\| \text{vec}\left( \mathbf{Y} \! - \! \mathbf{U}{{\mathbf{F}}^{\text{2D}}}{{\mathbf{X}}^{\left( k \right)}} \right) \right\|_{2}^{2} \! + \! \frac{{{\lambda }_{1}}}{2} \! \left\| \text{vec}\left( {{\mathbf{X}}^{\left( k \right)}} \! - \! \mathbf{G}{{\mathbf{X}}^{\left( k \right)}} \right) \right\|_{2}^{2},\\
 \end{aligned}
 \end{medsize}
\end{equation}

\begin{equation} \label{(SHLR-SV_Zrow)}
\begin{medsize}
\begin{aligned}
\mathbf{Z}{{_{m}^{\text{row}}}^{\left( k+1 \right)}} & = \underset{\mathbf{Z}_{m}^{\text{row}}}{\mathop{\min }}\,{{\left\| \mathbf{Z}{{_{m}^{\text{row}}}^{\left( k \right)}} \right\|}_{*}} \\ 
& + \frac{\beta }{2}\left\| {{{\mathbf{\tilde{H}}}}_{\text{vc}}}\mathbf{\tilde{W}}{{{\mathbf{\tilde{F}}}}^{\text{1D}}}{{\mathbf{P}}_{m}}{{\mathbf{X}}^{\left( k+1 \right)}} \! - \! \mathbf{Z}{{_{m}^{\text{row}}}^{\left( k \right)}} \! + \! \frac{\mathbf{D}{{_{m}^{\text{row}}}^{\left( k \right)}}}{\beta } \right\|_{F}^{2},
\end{aligned}
\end{medsize}
\end{equation}

\begin{equation} \label{(SHLR-SV_Zcol)}
\begin{medsize}
\begin{aligned}
\mathbf{Z}{{_{n}^{\text{col}}}^{\left( k+1 \right)}} & = \underset{\mathbf{Z}_{n}^{\text{col}}}{\mathop{\min }}\,{{\left\| \mathbf{Z}{{_{n}^{\text{col}}}^{\left( k \right)}} \right\|}_{*}}\\
 & +\frac{\beta }{2}\left\| {{{\mathbf{\tilde{H}}}}_{\text{vc}}}\mathbf{\tilde{W}}{{{\mathbf{\tilde{F}}}}^{\text{1D}}}{{\mathbf{Q}}_{n}}{{\mathbf{X}}^{\left( k+1 \right)}}-\mathbf{Z}{{_{n}^{\text{col}}}^{\left( k \right)}}+\frac{\mathbf{D}{{_{n}^{\text{col}}}^{\left( k \right)}}}{\beta } \right\|_{F}^{2},
\end{aligned}
\end{medsize}
\end{equation}

\begin{equation} \label{(SHLR-SV_Drow)}
\begin{medsize}
%\begin{aligned}
\mathbf{D}{{_{m}^{\text{row}}}^{\left( k+1 \right)}} \!=\! \mathbf{D}{{_{m}^{\text{row}}}^{\left( k \right)}} \!+\! \tau \! \left( {{{\mathbf{\tilde{H}}}}_{\text{vc}}}\mathbf{\tilde{W}}{{{\mathbf{\tilde{F}}}}^{1\text{D}}}{{\mathbf{P}}_{m}}{{\mathbf{X}}^{\left( k+1 \right)}}-\mathbf{Z}{{_{m}^{\text{row}}}^{\left( k+1 \right)}} \! \right),
%\end{aligned}
\end{medsize}
\end{equation}

\begin{equation} \label{(SHLR-SV_Dcol)}
\begin{medsize}
%\begin{aligned}
\mathbf{D}{{_{n}^{\text{col}}}^{\left( k+1 \right)}}=\mathbf{D}{{_{n}^{\text{col}}}^{\left( k \right)}}+\tau \left( {{{\mathbf{\tilde{H}}}}_{\text{vc}}}\mathbf{\tilde{W}}{{{\mathbf{\tilde{F}}}}^{1\text{D}}}{{\mathbf{Q}}_{n}}{{\mathbf{X}}^{\left( k+1 \right)}}-\mathbf{Z}{{_{n}^{\text{col}}}^{\left( k+1 \right)}} \right),
%\end{aligned}
\end{medsize}
\end{equation}
where $\tau$ is the step size.

For fixed $\mathbf{Z}{{_{m}^{\text{row}}}^{\left( k \right)}}$, $\mathbf{Z}{{_{n}^{\text{col}}}^{\left( k \right)}}$, $\mathbf{D}{{_{m}^{\text{row}}}^{\left( k \right)}}$, and $\mathbf{D}{{_{n}^{\text{col}}}^{\left( k \right)}}$, ${{\mathbf{X}}^{\left( k+1 \right)}}$ has a close-form solution as

\begin{equation} \label{(SHLR-SV_X_solution)}
\begin{medsize}
\begin{aligned}
  & {\mathbf{X}}^{\left( k+1 \right)} \!\! = \! \left( \! \lambda {{\mathbf{F}}^{\text{2D,*}}}{{\mathbf{U}}^{*}}\mathbf{U}{{\mathbf{F}}^{\text{2D}}} \!\! +\! \beta \!\! \sum\limits_{m=1}^{M} \!\! {\mathbf{P}_{m}^{*}{{{\mathbf{\tilde{F}}}}^{1\text{D,*}}}{{{\mathbf{\tilde{W}}}}^{*}}\mathbf{\tilde{H}}_{\text{vc}}^{*}{{{\mathbf{\tilde{H}}}}_{\text{vc}}}\mathbf{\tilde{W}}{{{\mathbf{\tilde{F}}}}^{1\text{D}}}{{\mathbf{P}}_{m}}} \right.\\
  & + {{\left. {{\lambda }_{1}}{{\left( \mathbf{G} \! - \! \mathbf{I} \right)}^{*}} \!\! \left( \mathbf{G} \! - \! \mathbf{I} \right) \! + \!\beta \! \sum\limits_{n=1}^{N}{\mathbf{Q}_{n}^{*}{{{\mathbf{\tilde{F}}}}^{1\text{D,*}}}{{{\mathbf{\tilde{W}}}}^{*}}\mathbf{\tilde{H}}_{\text{vc}}^{*}{{{\mathbf{\tilde{H}}}}_{\text{vc}}}\mathbf{\tilde{W}}{{{\mathbf{\tilde{F}}}}^{1\text{D}}}{{\mathbf{Q}}_{n}}} \right)}^{-1}}  \\ 
  & \left ( \lambda {{\mathbf{F}}^{\text{2D,*}}}{{\mathbf{U}}^{*}}\mathbf{Y} \!+\!\beta \!\! \sum\limits_{m=1}^{M}{\mathbf{P}_{m}^{*}{{{\mathbf{\tilde{F}}}}^{1\text{D,*}}}{{{\mathbf{\tilde{W}}}}^{*}}\mathbf{\tilde{H}}_{\text{vc}}^{*}\left( \mathbf{Z}{{_{m}^{\text{row}}}^{\left( k \right)}}-\frac{\mathbf{D}{{_{m}^{\text{row}}}^{\left( k \right)}}}{\beta } \right)} \right . \\
 & \left . + \beta \sum\limits_{n=1}^{N}{\mathbf{Q}_{n}^{*}{{{\mathbf{\tilde{F}}}}^{1\text{D,*}}}{{{\mathbf{\tilde{W}}}}^{*}}\mathbf{\tilde{H}}_{\text{vc}}^{*}\left( \mathbf{Z}{{_{n}^{\text{col}}}^{\left( k \right)}}-\frac{\mathbf{D}{{_{n}^{\text{col}}}^{\left( k \right)}}}{\beta } \right)} \right ),
\end{aligned}
\end{medsize}
\end{equation}
where the upper subscript $*$ denotes the adjoint operator. Here, for any matrix $\mathbf{A}$, $\text{vec}^{-1}\left( \text{vec} \left( \mathbf{A} \right) \right) = \mathbf{I}$, therefore, we omitted the writing of $\text{vec}^{-1}\left( \text{vec} \left( \cdot \right) \right)$ for simplicity in the following.

For fixed ${{\mathbf{X}}^{\left( k+1 \right)}}$ and $\mathbf{D}{{_{m}^{\text{row}}}^{\left( k \right)}}$, $\mathbf{Z}{{_{m}^{\text{row}}}^{\left( k+1 \right)}}$ is obtained by
\begin{equation} \label{(SHLR-SV_Zrow_solution)}
\begin{medsize}
\mathbf{Z}{{_{m}^{\text{row}}}^{\left( k+1 \right)}}={{S}_{1/\beta }}\left( {{{\mathbf{\tilde{H}}}}_{\text{vc}}}\mathbf{\tilde{W}}{{{\mathbf{\tilde{F}}}}^{1\text{D}}}{{\mathbf{P}}_{m}}{{\mathbf{X}}^{\left( k+1 \right)}}+\frac{\mathbf{D}{{_{m}^{\text{row}}}^{\left( k \right)}}}{\beta } \right),
\end{medsize}
\end{equation}
where ${{S}_{1/\beta }}\left( \cdot  \right)$ denotes the singular value threshold operator with the threshold of $1/\beta $. 

Similarly, the $\mathbf{Z}{{_{n}^{\text{col}}}^{\left( k+1 \right)}}$ can also be obtained by
\begin{equation} \label{(SHLR-SV_Zcol_solution)}
\begin{medsize}
\mathbf{Z}{{_{n}^{\text{col}}}^{\left( k+1 \right)}}={{S}_{\frac{1}{\beta }}}\left( {{{\mathbf{\tilde{H}}}}_{\text{vc}}}\mathbf{\tilde{W}}{{{\mathbf{\tilde{F}}}}^{1\text{D}}}{{\mathbf{Q}}_{n}}{{\mathbf{X}}^{\left( k+1 \right)}}+\frac{\mathbf{D}{{_{n}^{\text{col}}}^{\left( k \right)}}}{\beta } \right).
\end{medsize}
\end{equation}

The numerical algorithm for SHLR-SV is summarized in Supplementary Material S1.

\subsubsection{SHLR-VP}
Considering the space limitations, please refer to Supplementary Material S2 for detailed derivations for solving the SHLR-VP model.

%----------------------------------------------------------------------
%------------------------------- Results ------------------------------
%----------------------------------------------------------------------
\section{Results}\label{Section:results}

\begin{figure*}[htbp]
\setlength{\abovecaptionskip}{0.cm}
\setlength{\belowcaptionskip}{-0.cm}
\centering
\includegraphics[width=6.4in]{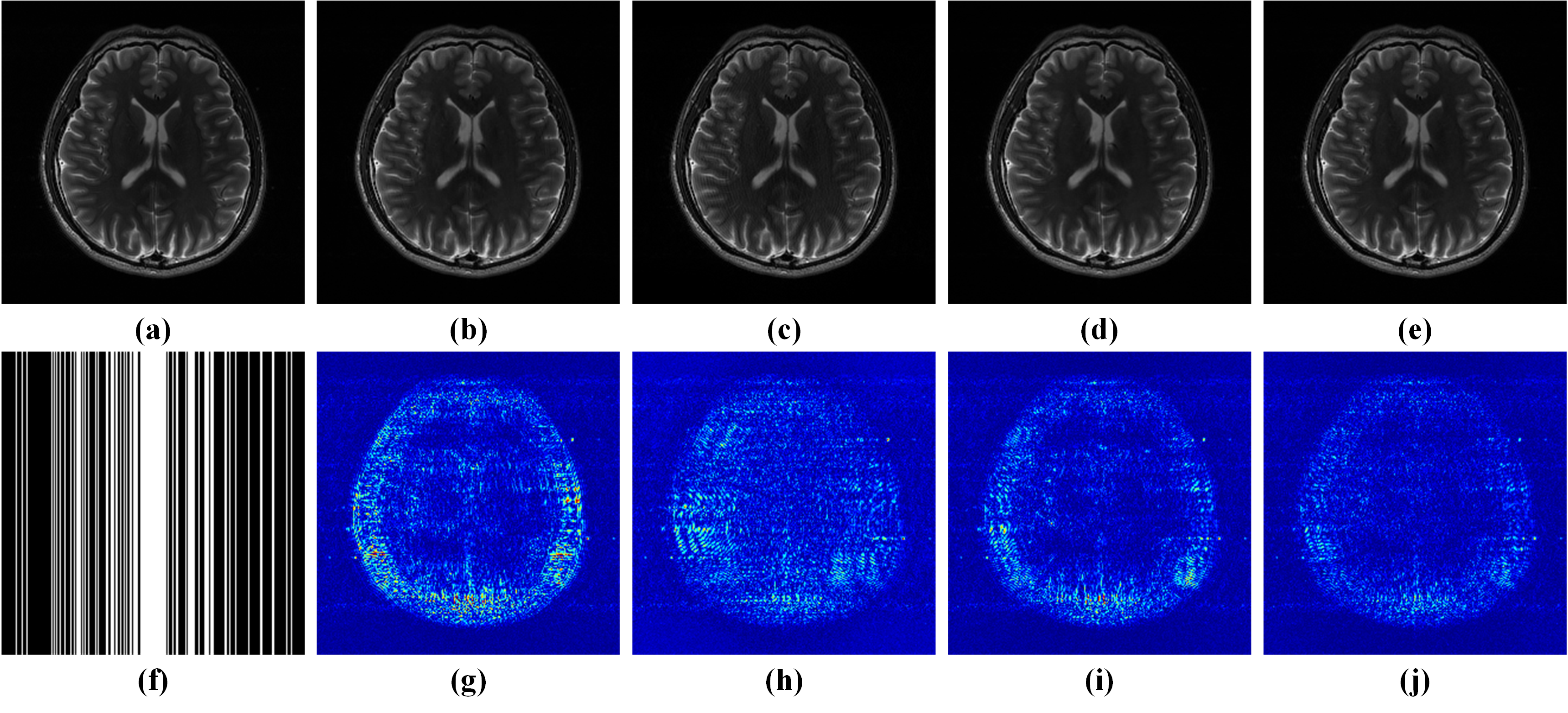}
\caption{Parallel imaging reconstruction results and errors under two patterns. (a) An SSOS image of fully sampled data; (b-e) SSOS images of reconstructed results by $\ell_1$-SPIRiT, AC-LORAKS, STDLR-SPIRiT and SHLR-SV, respectively; (f) the Cartesian undersampling pattern with a sampling rate of 0.34; (g-j) the reconstruction error distribution (12.5$\times$) corresponding to reconstructed image above them.}
\label{fig_2D_random}
\end{figure*}

In this section, we will evaluate the two proposed approaches, SHLR-SV and SHLR-VP using \textit{in vivo} data from different MRI scanners. 

In all experiments, the multi-coil images are combined by a square root of squares (SSOS). We adopted the RLNE \cite{2021_MIA_xinlin} and mean structure similarity index measure (MSSIM) \cite{2004_MSSIM} as objective criteria to quantify the reconstruction performance. A lower RLNE, the higher consistency lies between the fully sampled image and the reconstruction image. A higher MSSIM indicates higher detail preservation in reconstruction. The calculation of RLNE and MSSIM can be found in Supplementary Material.

\subsection{Parallel Imaging}
In this subsection, we assessed the reconstruction results and running time of SHLR-SV and compared the results of SHLR-SV with three state-of-the-art reconstruction approaches including $\ell_1$-SPIRiT \cite{2010_SPIRiT}, AC-LORAKS \cite{2015_AC-LORAKS}, and STDLR-SPIRiT \cite{2020_xinlin}. These three methods and SHLR-SV require ACS for reconstruction. More specifically, $\ell_1$-SPIRiT reconstructs image by exploiting kernel estimation and the sparsity of the image in the transform domain. Both AC-LORAKS and STDLR-SPIRiT utilize the low-rankness of block Hankel matrix (or structured matrix) of k-space to recover MRI image. In addition, AC-LORAKS utilizes ACS for reducing the algorithm complexity \cite{2015_AC-LORAKS}.

The codes of $\ell_1$-SPIRiT are shared online by Dr. Michael Lustig \cite{code_SPIRiT} and codes of AC-LORAKS \cite{code_LORAKS} are shared at Dr. Justin P. Haldar's website. Here, we adopted the S-based AC-LORAKS with virtual coil which makes use of phase constraints. The reason why we chose S-based is that LORAKS with phase constraint provides results with lower error compared to other constraints \cite{2014_LORAKS}. Parameters of all the compared methods are optimized to obtain the lowest RLNE. For the proposed methods, the effect of parameters setting will be discussed in Supplementary Material S5.

Two brain datasets acquired from healthy volunteers are adopted in experiments. The dataset depicted in Fig. \ref{fig_2D_uniform} (a) is obtained from a 3T SIEMENS MRI scanner (Siemens Healthcare, Erlangen, Germany) equipped with 32 coils using T2-weighted turbo spin echo sequence (matrix size = $256 \times 256$, TR/TE = $3000 / 66$ ms, FOV = $200$ mm $\times$ $200$ mm, slice thickness = $5$ mm). Eight virtual coils are compressed from the acquired data of 32 coils \cite{2013_Coil_Compression} to reduce the computational complexity. The other dataset shown in Fig. \ref{fig_2D_random} (a) is acquired from a 3T SIEMENS Trio whole-body scanner (Siemens Healthcare, Erlangen, Germany) equipped with  32 coils using the 2D T2-weighted turbo spin echo sequence (matrix size = $256 \times 256$, TR/TE = $6100 / 99$ ms, FOV = $220$ mm $\times$ $220$ mm, slice thickness = $3$ mm). Four virtual coils are compressed from the acquired data of 32 coils \cite{2013_Coil_Compression}.

We first test the proposed method for parallel imaging using the widely-used sampling pattern adopted in commercial MRI scanners named uniform undersampling pattern. Both $\ell_1$-SPIRiT (Fig. \ref{fig_2D_uniform} (b)) and AC-LORAKS (Fig. \ref{fig_2D_uniform} (c)) have strong undersampling artifacts. STDLR-SPIRiT (Fig. \ref{fig_2D_uniform} (d)) and SHLR-SV (Fig. \ref{fig_2D_uniform} (e)) show the good ability of artifacts removing. But it is worthy to note that SHLR-SV provides the image with lower error than STDLR-SPIRiT.

We then test the proposed method using non-uniform undersampling patterns, including 1D Cartesian and 2D random undersampling patterns. In the reconstruction under Cartesian undersampling pattern, AC-LORAKS (Fig. \ref{fig_2D_random} (c)) yields results exhibiting obvious artifacts inside the brain area, whereas the ringing artifacts also remain in the reconstructed image of $\ell_1$-SPIRiT (Fig. \ref{fig_2D_random} (b)). Both STDLR-SPIRiT (Fig. \ref{fig_2D_random} (d)) and SHLR-SV (Fig. \ref{fig_2D_random} (e)) provide the image with nice artifacts suppression, however, the reconstruction error of STDLR-SPIRiT appear slightly larger than that of SHLR-SV, especially in the skull area.

\begin{figure}[!htb]
\setlength{\abovecaptionskip}{0.cm}
\setlength{\belowcaptionskip}{-0.cm}
\centering
\includegraphics[width=3.4in]{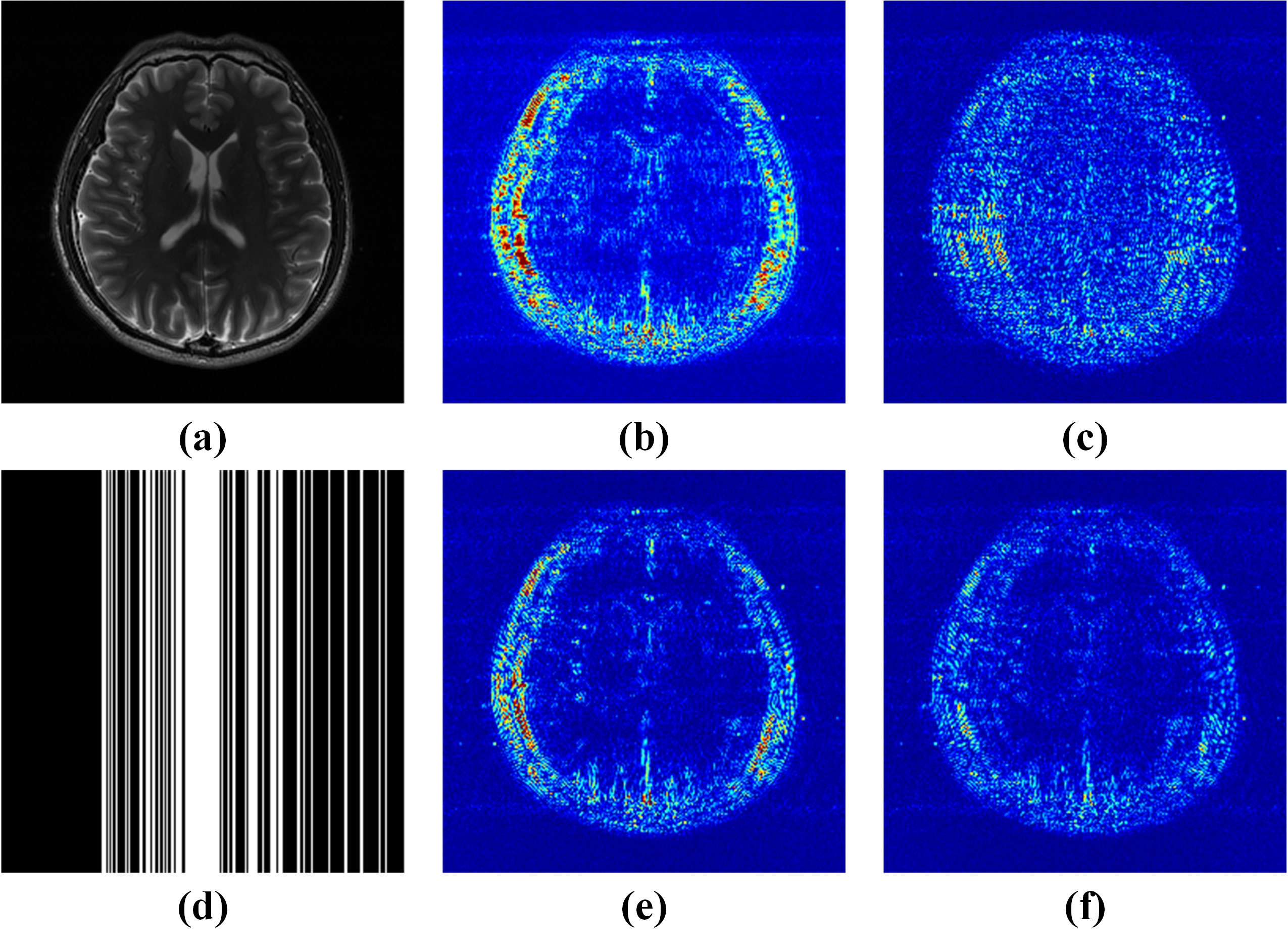}
\caption{Parallel imaging reconstruction resutls using the undersampling pattern with patrial Fourier. (a) An SSOS image of fully sampled data; (b, c, e, f) reconstruction errors (10$\times$) introduced by $\ell_1$-SPIRiT, AC-LORAKS, STDLR-SPIRiT and SHLR-SV, respectively; (d) the Cartesian undersampling pattern with 3/4 PF (total sampling rate 0.30).}
\label{fig_2D_PF}
\end{figure}

\begin{table}[htbp]
  \centering
  \caption{RLNE/MSSIM for parallel imaging reconstructions.}
    \begin{tabular}{m{0.7cm}<{\centering}m{1.35cm}<{\centering}m{1.6cm}<{\centering}m{1.85cm}<{\centering}m{1.25cm}<{\centering}}
    % \begin{tabular}{ccccc}
    \toprule
    Image & $\ell_1$-SPIRiT & AC-LORAKS & STDLR-SPIRiT & SHLR-SV \\
    \midrule
    Fig. \ref{fig_2D_uniform} & 0.0688/0.9637 & 0.0709/0.9386 & 0.0488/0.9810 & \textbf{0.0482}/\textbf{0.9822} \\
    Fig. \ref{fig_2D_random} & 0.0689/0.9856 & 0.0596/0.9806 & 0.0518/0.9918 & \textbf{0.0426}/\textbf{0.9934} \\    
    Fig. \ref{fig_2D_PF} & 0.1028/0.9751 & 0.0935/0.9543 & 0.0914/0.9833 & \textbf{0.0639}/\textbf{0.9882} \\
    \bottomrule
    \end{tabular}
  \label{Table_2D}%
\end{table}%

In addition, SHLR-SV allows promising reconstruction on the undersampling patterns with partial Fourier (PF), which is commonly adopted in commercial equipment for further acceleration \cite{1991_PF}, indicating the potential to be robust to patterns. As shown in Fig. \ref{fig_2D_PF}, under the case with PF, the reconstruction errors of three compared approaches, including STDLR-SPIRiT, apparently increased, whereas SHLR-SV still yields nice reconstruction with low reconstruction error. In the meanwhile, the proposed SHLR-SV outperforms other methods in terms of RLNE and MSSIM, indicating the excellent ability of artifact suppression and detail preservation.

We include more data and sampling patterns to further evaluate the proposed SHLR-SV approach. The detailed results can be found in Supplementary Material S3. Similar phenomenons were observed in the results indicating SHLR-SV is robust to the different data from different subjects, and also robust to different sampling patterns. It can be seen that SHLR-SV outperforms the $\ell_1$-SPIRiT and AC-LORAKS, and shows comparable results with STDLR-SPIRiT. Given the patterns with PF, SHLR-SV gains a slight improvement than STDLR-SPIRiT with respect to RLNE and MSSIM. However, the computational time of the proposed method is significantly reduced compared to that of STDLR-SPIRiT ($8 \times$ faster). Overall, the proposed method provides faithful reconstruction with acceptable computational time, having the potential to be used in clinical applications.
 
 \begin{figure*}[htbp]
\setlength{\abovecaptionskip}{0.cm}
\setlength{\belowcaptionskip}{-0.cm}
\centering
\includegraphics[width=5.6in]{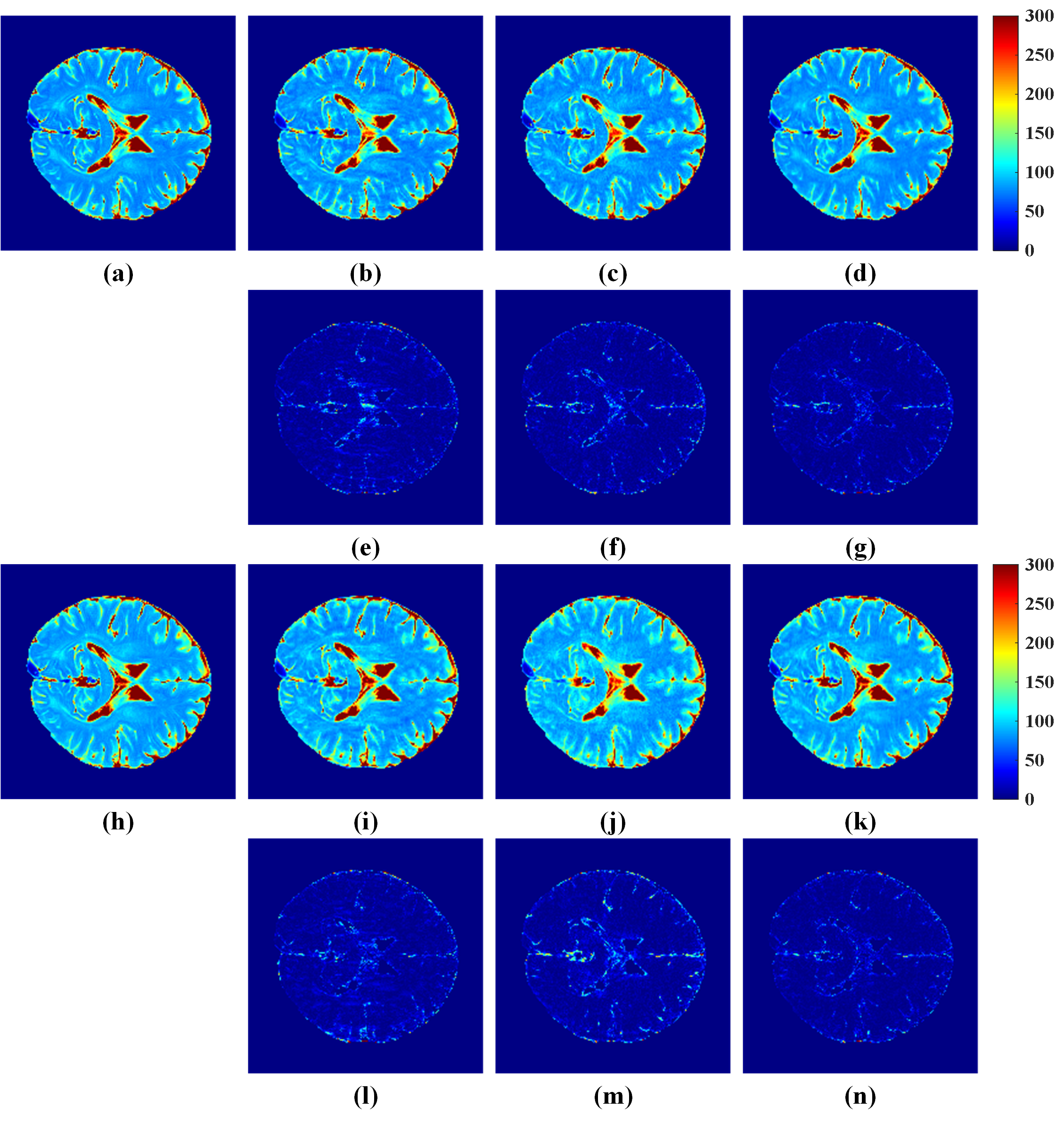}
\caption{T2 mapping reconstruction results and errors under two patterns. (a) (or (h)) T2 map of fully sampled data; (b-d) T2 map images of reconstructed results at reduction factors R=6 by MORASA, ALOHA, and SHLR-VP, respectively; (i-k) T2 maps of reconstructed results at R=8 by MORASA, ALOHA, and SHLR-VP, respectively; (e-g) and (l-n) the reconstruction error distribution (6$\times$) corresponding to reconstructed image above them. Note: the RLNE/MSSIM of (b-d) are 0.0978/0.9891, 0.0959/0.9883, and \textbf{0.0882}/\textbf{0.9909}, respectively, and the RLNE/MSSIM of (i-k) are 0.1262/0.9831, 0.1386/0.9775, and \textbf{0.1029}/\textbf{0.9872}, respectively.}
\label{fig_mapping}
\end{figure*}

\subsection{Parameter Imaging}
In this subsection, we validate the proposed SHLR-VP method using a acquired \textit{in-vivo} T2 mapping data.
We compare the proposed method named SHLR-VP with two state-of-the-art mapping reconstruction methods including MORASA \cite{2016_MORASA}, and ALOHA \cite{2016_ALOHA_mapping}. These two methods both utilized the property of exponential in the parameter dimension. MORASA modeled the time-domain signal into a exponential function and the enforced the sparsity of the image at each time point. ALOHA made use of the annihilation relationship of the data and lifted the signal into a higher dimensional space to ensure the low-rankness. The codes of MORASA are shared by Dr. Xi Peng and codes of ALOHA \cite{code_ALOHA} are shared at Dr. Jong Chul Ye website.

A fully sampled brain dataset was acquired from a healthy volunteer on a 3T 12 coils MRI scanner (Siemens Healthcare, Erlangen, Germany) using turbo spin echo sequence (TR = $4000$ ms, $15$ TEs from $8.8$ to $132$ ms with $8.8$ ms sapcing, FOV = $200$ mm $\times$ $200$ mm, matrix size = $192 \times 192$, slice thickness = $3$ mm). To estimate the T2 map, standard nonlinear least square fitting method was performed in the selected region-of-interest on a pixel-by-pixel basis. T2 relaxation values outside the reasonable range were excluded from the T2 map.

We performed reconstructions under two different reduction factors, $R = 6$ and $R = 8$. In both cases, the MORASA images appear apparent string-like artifacts (Fig. \ref{fig_mapping} (e) and (i)). Compared to ALOHA, SHLR-VP produces mapping results with better accuracy. As can be seen from the images, the mapping error is obviously smaller than that of ALOHA. The RLNE and MSSIM also indicate that SHLR-VP persevere the image structure better than the comparison methods.

In addition, we validated the proposed method using other datasets, and the sampling patterns with PF. Please refer to the Supplementary Material S4 for detailed results. The results demonstrate that the proposed method SHLR-VP yields reconstructions with the lowest error in both reconstructed images and the quantization T2 map.

%----------------------------------------------------------------------
%---------------------------- Discussion ------------------------------
%----------------------------------------------------------------------
\section{Discussions}\label{Section:discussion}
\subsection{The Number of ACS}
The ACS is crucial for traditional parallel reconstruction methods. These methods estimate sensitivity maps or kernels from ACS, once the number of ACS lines is limited, the estimation turns inaccurate, leading to failures of reconstruction \cite{2014_MRM_Lustig}. Four comparison methods for parallel imaging require ACS to reconstruct images, thus, in this subsection we discuss the effect of the number of ACS on the reconstructed results. 

\begin{figure}[htbp]
\setlength{\abovecaptionskip}{0.cm}
\setlength{\belowcaptionskip}{-0.cm}
\centering
\includegraphics[width=3.4in]{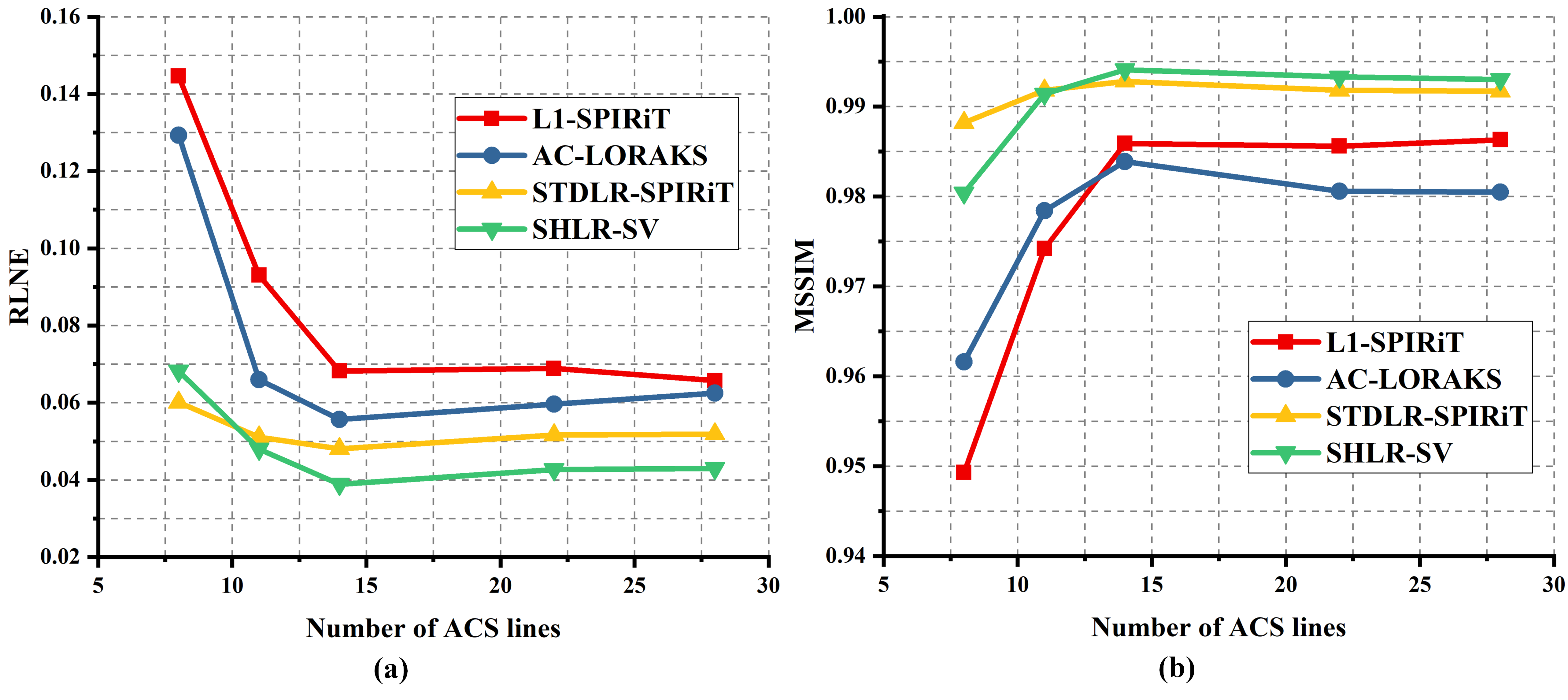}
\caption{The variation trends of RLNE (a) and MSSIM (b) versus the number of ACS lines under parallel imaging reconstruction. This experiment is carried out with a series of Gaussian Cartesian sampling patterns but all at the same sampling rate of 0.34. The difference between each sampling pattern is the number of ACS lines. Note: all experiments are conducted on the brain data shown in Fig. \ref{fig_2D_random} (a). All methods are performed with the same parameters respectively, with which these methods provide the lowest RLNE when ACS lines equal 22.}
\label{fig_ACS}
\end{figure}

The results shown in Fig. \ref{fig_ACS} indicates the proposed SHLR-SV maintains the robustness to ACS lines. When ACS lines ($>14$ lines) are relatively sufficient, all these comparison methods provide reconstruction with relatively low reconstruction error. Especially, the proposed method SHLR-SV achieved the lowest RLNE and the highest MSSIM. However, when the number of ACS lines is extremely low ($=8$), $\ell_1$-SPIRiT and AC-LORAKS fail to produce satisfying results (the RLNE reaches closely to $0.2$). In contrast, the STDLR-SPIRiT and SHLR-SV, still work well, even with only $8$ ACS lines offering an RLNE of approximately 0.07. The observation demonstrates that the proposed SHLR-SV still inherits the robustness of structured low-rank methods.

\subsection{Computational Time}
All reconstruction software were ran on a CentOS 7 computation server with two Intel Xeon CPUs of 3.5 GHz and 112 GB RAM. The computational time is obtained by averaging the reconstruction time of $10$ Monte Carlo tests. All comparison approaches were implemented in MATLAB.

\begin{figure}[htbp]
\setlength{\abovecaptionskip}{0.cm}
\setlength{\belowcaptionskip}{-0.cm}
\centering
\includegraphics[width=3.2in]{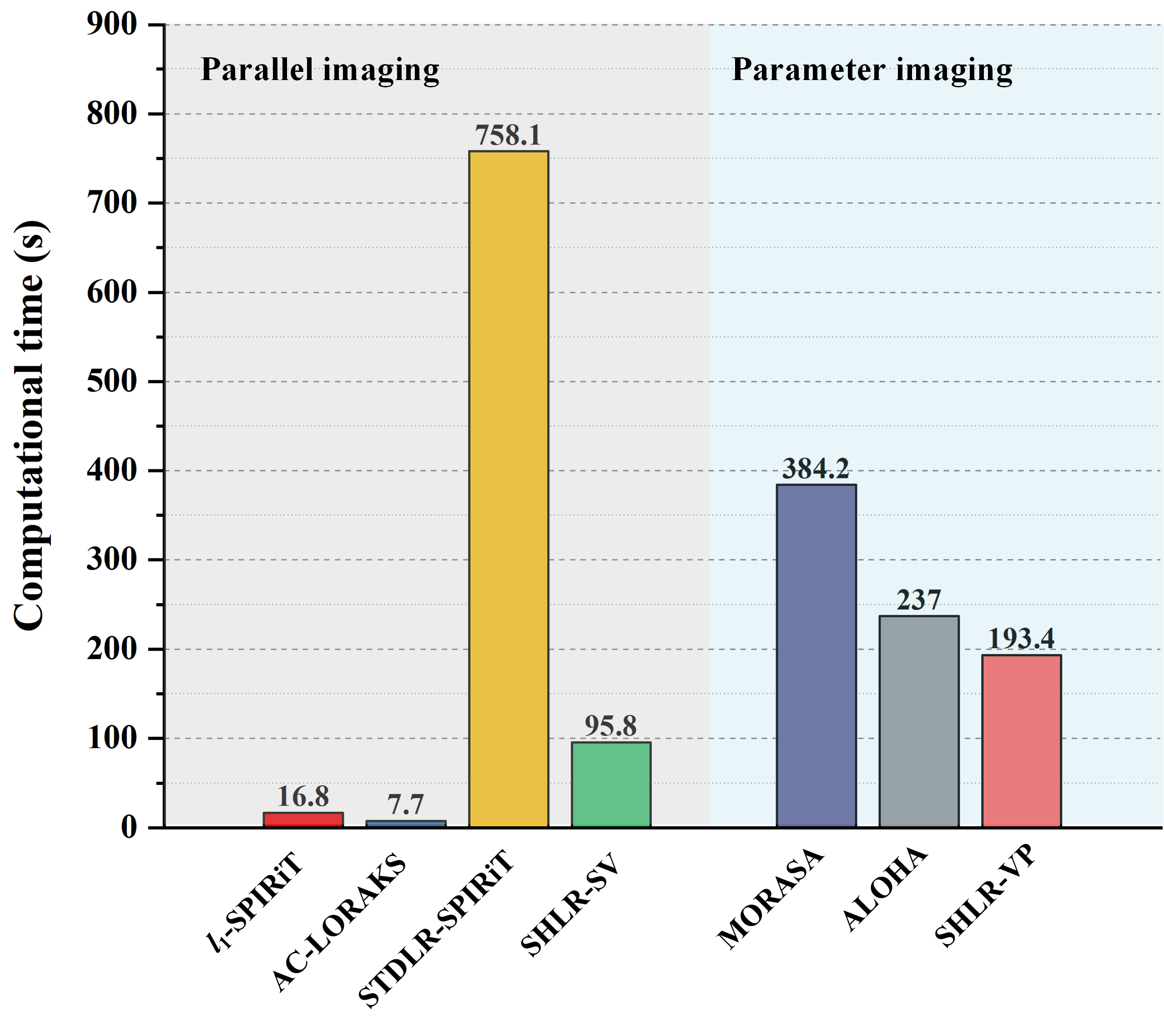}
\caption{The computational time of different methods in spatial imaging and parameter imaging. Note: parallel imaging reconstrctions are conducted on the brain data shown in Fig. \ref{fig_2D_random} (a) under Gaussian Cartesian undersampling pattern shown in Fig. \ref{fig_2D_random} (f), and parameter imaging reconstructions are conducted on the data shown in Fig. \ref{fig_mapping} with the reduction factor R=8.}
\label{fig_runtime}
\end{figure}

As shown in Fig. \ref{fig_runtime}, the running time of SHLR-SV is reduced to $1/8$ of that of STDLR-SPIRiT, which substantially alleviates the burden of lengthy reconstruction time of STDLR-SPIRiT. Moreover, it can be observed that the proposed method runs relatively slower than $\ell_1$-SPIRiT and AC-LORAKS. However, the additional computational cost of SHLR-SV should be acceptable considering its improvement in image reconstruction.

The computational time on parameter imaging reconstruction is also depicted in Fig. \ref{fig_runtime}. It shown that the proposed SHLR-VP reconstructs data at a faster speed than MORASA and ALOHA. Besides, it is worthy to notice that runtime of the proposed method can be further reduced with GPU or MEX/C implementations.

%----------------------------------------------------------------------
%----------------------------- Conclusion -----------------------------
%----------------------------------------------------------------------
\section{Conclusion}\label{Section:conclusion}
In this work, we attempted to alleviate the huge computational complexity and lengthy reconstruction time problems of the state-of-the-art structural low-rank approaches. To reduce the computational time, we proposed a separable Hankel low-rank reconstruction method, named SHLR, to enforce the Hankel low-rankness of each row and each column of the signal of interest. But this way to enforce the low-rankness is sub-optimal compared to block Hankel methods, though holds the advantage of swift runtime. To enhance the performance of SHLR, we introduced prior information. For parallel imaging, we explored the correlation between rows/columns and the virtual coil technique, and proposed the SHLR-SV model. Regarding parameter imaging, we assumed that the signal intensity along the parameter dimension varies in an exponential way. We enforced the exponential property and virtual coil technique and proposed a model named SHLR-VP for accelerating parameter imaging.

Experimental \textit{in-vivo} results showed that the proposed two approaches enable better results than the state-of-the-art methods. Notably, the proposed methods allow reconstructions with smaller errors and faster reconstruction speed.

%----------------------------------------------------------------------
%----------------------------- References -----------------------------
%----------------------------------------------------------------------
% \ifCLASSOPTIONcaptionsoff
%  \newpage
% \fi

% \newpage
\bibliographystyle{IEEEtran}
\bibliography{IEEEabrv,Mylib}

\end{document}

\title{Supplementary Material of ‘‘Accelerated MRI Reconstruction with Separable and Enhanced Low-Rank Hankel Regularization’’}

\author{Xinlin Zhang, Hengfa Lu, Di Guo, Zongying Lai, Huihui Ye, Xi Peng, Bo Zhao, and Xiaobo Qu}

\maketitle

\IEEEpeerreviewmaketitle
\renewcommand\thefigure{S\arabic{figure}}{}
\renewcommand\theequation{S\arabic{equation}}{}{}
\renewcommand\thesection{S\arabic{section}}{}
\renewcommand\thealgorithm{S\arabic{algorithm}}{}
\renewcommand\thetable{S\arabic{table}}{}

%----------------------------------------------------------------------
%---------------------------- Supplement ------------------------------
%----------------------------------------------------------------------

\section{Numerical algorithm for SHLR-SV}

The numerical algorithm for SHLR-SV is summarized in Algorithm \ref{alg:1}.

\begin{algorithm}[htb]
\footnotesize
\caption{MRI reconstruction with SHLR-SV.}\label{alg:1}
\hspace*{0.02in}{\bf{Input:}}  $\mathbf{Y}$, $\mathbf{U}$, $\mathbf{G}$, $\lambda $, ${{\lambda }_{1}}$, $\beta $, $\tau $.\\
\hspace*{0.02in}{\bf{Initialization:}} $\mathbf{Z}_{m}^{\text{row}}=\mathbf{Z}_{n}^{\text{col}}=\mathbf{0}$, $\mathbf{D}_{m}^{\text{row}}=\mathbf{D}_{n}^{\text{col}}=\mathbf{1}$, and $k=1$.\\
\hspace*{0.02in}{\bf{Output:}} ${\mathbf{X}}$.
\begin{algorithmic}[1]
\WHILE{ $k\le 50$ and $  \left\| {{\mathbf{X}}^{\left( k+1 \right)}}-{{\mathbf{X}}^{\left( k \right)}}\  \right\|_{F}^{2}\ /\left\| {{\mathbf{X}}^{\left( k \right)}}\  \right\|_{F}^{2}\ge {{10}^{-6}}$ }
\STATE
Update ${{\mathbf{X}}^{\left( k+1 \right)}}$ by solving Eq. (15);
\STATE
Update $\mathbf{Z}{{_{m}^{\text{row}}}^{\left( k+1 \right)}}$ and $\mathbf{Z}{{_{n}^{\text{col}}}^{\left( k+1 \right)}}$ by using Eq. (16) and Eq. (17) in the main text;
\STATE
Update multiplier $\mathbf{D}{{_{m}^{\text{row}}}^{\left( k+1 \right)}}$ and $\mathbf{D}{{_{n}^{\text{col}}}^{\left( k+1 \right)}}$ by using Eq. (13) and Eq. (14) in the main text;
\STATE $ k = k + 1 $;
\ENDWHILE
\end{algorithmic}
\end{algorithm}

\section{Derivations and numerical algorithm for SHLR-VP}

The augmented Lagrangian form of Eq. (8) in the main text is
\begin{equation} \label{(SHLR-P_ALF)}
\begin{medsize}
\begin{aligned}
 \underset{\mathbf{D}_{n}^{\text{PE}},\mathbf{D}_{l}^{\text{P}}}{\mathop{\max }}\,\underset{\mathbf{X}_{m}^{\text{PE-P}},\mathbf{Z}_{n}^{\text{PE}},\mathbf{Z}_{l}^{\text{P}}}{\mathop{\min }} & \frac{\lambda }{2}\left\| \text{vec}\left( \mathbf{Y}_{m}^{\text{PE-P}}-\mathbf{U}{{\mathbf{F}}^{\text{PE}}}\mathbf{X}_{m}^{\text{PE-P}} \right) \right\|_{2}^{2} + 
 \sum\limits_{n=1}^{N}{\left( {{\lambda }_{2}}{{\left\| \mathbf{Z}_{n}^{\text{PE}} \right\|}_{*}} \! + \! \left\langle \mathbf{D}_{n}^{\text{PE}},\mathbf{\tilde{H}}{{\mathbf{P}}_{n}}\mathbf{X}_{m}^{\text{PE-P}} \right\rangle \! + \! \frac{\beta }{2}\left\| \mathbf{\tilde{H}}{{\mathbf{P}}_{n}}\mathbf{X}_{m}^{\text{PE-P}} \! - \! \mathbf{Z}_{n}^{\text{PE}} \right\|_{F}^{2} \right)} \\ 
 & + \sum\limits_{l=1}^{L}{\left( {{\left\| \mathbf{Z}_{l}^{\text{P}} \right\|}_{*}} \! + \! \left\langle \mathbf{D}_{l}^{\text{P}},{{{\mathbf{\tilde{H}}}}_{\text{vc}}}\mathbf{\tilde{W}}{{{\mathbf{\tilde{F}}}}^{1\text{D}}}{{\mathbf{Q}}_{l}}\mathbf{X}_{m}^{\text{PE-P}} \right\rangle \! + \! \frac{\beta }{2}\left\| {{{\mathbf{\tilde{H}}}}_{\text{vc}}}\mathbf{\tilde{W}}{{{\mathbf{\tilde{F}}}}^{1\text{D}}}{{\mathbf{Q}}_{l}}\mathbf{X}_{m}^{\text{PE-P}} \!- \! \mathbf{Z}_{l}^{\text{P}} \right\|_{F}^{2} \right)}.
\end{aligned}
\end{medsize}
\end{equation}

The solution of Eq. \eqref{(SHLR-P_ALF)} is obtained by alternatively solving the following sub-problems:

\begin{equation} \label{(SHLR-P_X)}
\begin{medsize}
\begin{aligned}
\mathbf{X}{{_{m}^{\text{PE-P}}}^{\left( k+1 \right)}} = & \arg \underset{\mathbf{X}_{m}^{\text{PE-P}}}{\mathop{\min }}\,  \frac{\lambda }{2} \left\| \text{vec}\left( \mathbf{Y}_{m}^{\text{PE-P}}-\mathbf{U}{{\mathbf{F}}^{\text{PE}}}\mathbf{X}{{_{m}^{\text{PE-P}}}^{\left( k \right)}} \right) \right\|_{2}^{2} 
 +\sum\limits_{n=1}^{N}{\frac{\beta }{2}\left\| \mathbf{\tilde{H}}{{\mathbf{P}}_{n}}\mathbf{X}{{_{m}^{\text{PE-P}}}^{\left( k \right)}}-\mathbf{Z}{{_{n}^{\text{PE}}}^{\left( k \right)}}+\frac{\mathbf{D}{{_{n}^{\text{PE}}}^{\left( k \right)}}}{\beta} \right\|_{F}^{2}} \\
 & +\sum\limits_{l=1}^{L}{\frac{\beta }{2}\left\| {{{\mathbf{\tilde{H}}}}_{\text{vc}}}\mathbf{\tilde{W}}{{{\mathbf{\tilde{F}}}}^{1\text{D}}}{{\mathbf{Q}}_{l}}\mathbf{X}{{_{m}^{\text{PE-P}}}^{\left( k \right)}}-\mathbf{Z}{{_{l}^{\text{P}}}^{\left( k \right)}}+\frac{\mathbf{D}{{_{l}^{\text{P}}}^{\left( k \right)}}}{\beta } \right\|_{F}^{2}},
\end{aligned}
\end{medsize}
\end{equation}

\begin{equation} \label{(SHLR-P_Zpe)}
\mathbf{Z}{{_{n}^{\text{PE}}}^{\left( k+1 \right)}} = \arg \underset{\mathbf{Z}_{n}^{\text{PE}}}{\mathop{\min }}\,{{\lambda }_{2}}{{\left\| \mathbf{Z}{{_{n}^{\text{PE}}}^{\left( k \right)}} \right\|}_{*}} + \frac{\beta }{2}\left\| \mathbf{\tilde{H}}{{\mathbf{P}}_{n}}\mathbf{X}{{_{m}^{\text{PE-P}}}^{\left( k+1 \right)}}  -  \mathbf{Z}{{_{n}^{\text{PE}}}^{\left( k \right)}}  +  \frac{\mathbf{D}{{_{n}^{\text{PE}}}^{\left( k \right)}}}{\beta } \right\|_{F}^{2},
\end{equation}

\begin{equation} \label{(SHLR-P_Zt)}
\mathbf{Z}{{_{l}^{\text{P}}}^{\left( k+1 \right)}} = \arg \underset{\mathbf{Z}_{l}^{\text{P}}}{\mathop{\min }}\,{{\left\| \mathbf{Z}{{_{l}^{\text{P}}}^{\left( k \right)}} \right\|}_{*}} + \frac{\beta }{2}\left\| {{{\mathbf{\tilde{H}}}}_{\text{vc}}}\mathbf{\tilde{W}}{{{\mathbf{\tilde{F}}}}^{1\text{D}}}{{\mathbf{Q}}_{l}}\mathbf{X}{{_{m}^{\text{PE-P}}}^{\left( k+1 \right)}}  -  \mathbf{Z}{{_{l}^{\text{P}}}^{\left( k \right)}}  +  \frac{\mathbf{D}{{_{l}^{\text{P}}}^{\left( k \right)}}}{\beta } \right\|_{F}^{2},
\end{equation}

\begin{equation} \label{(SHLR-P_Dpe)}
\mathbf{D}{{_{n}^{\text{PE}}}^{\left( k+1 \right)}}=\mathbf{D}{{_{n}^{\text{PE}}}^{\left( k \right)}}+\tau \left( \mathbf{\tilde{H}}{{\mathbf{P}}_{n}}\mathbf{X}{{_{m}^{\text{PE-P}}}^{\left( k+1 \right)}}-\mathbf{Z}{{_{n}^{\text{PE}}}^{\left( k+1 \right)}} \right),
\end{equation}

\begin{equation} \label{(SHLR-P_Dt)}
\mathbf{D}{{_{l}^{\text{P}}}^{\left( k+1 \right)}}=\mathbf{D}{{_{l}^{\text{P}}}^{\left( k \right)}}+\tau \left( \mathbf{\tilde{H}\tilde{W}}{{{\mathbf{\tilde{F}}}}^{1\text{D}}}{{\mathbf{Q}}_{l}}\mathbf{X}{{_{m}^{\text{PE-P}}}^{\left( k+1 \right)}}-\mathbf{Z}{{_{l}^{\text{P}}}^{\left( k+1 \right)}} \right).
\end{equation}

The solution of Eq. \eqref{(SHLR-P_X)}

\begin{equation} \label{(SHLR-P_X_solution)}
\begin{medsize}
\begin{aligned}
 \mathbf{X}{{_{m}^{\text{PE-P}}}^{\left( k+1 \right)}} = &\left( \lambda {{\mathbf{F}}^{\text{PE,*}}}{{\mathbf{U}}^{*}}\mathbf{UF}^{\text{PE}} + \beta \sum\limits_{n=1}^{N}{\mathbf{P}_{n}^{*}{{{\mathbf{\tilde{H}}}}^{*}}\mathbf{\tilde{H}}{{\mathbf{P}}_{n}}} + \beta \sum\limits_{l=1}^{L}{\mathbf{Q}_{l}^{*}{{{\mathbf{\tilde{F}}}}^{1\text{D,*}}}{{{\mathbf{\tilde{W}}}}^{*}}{{{\mathbf{\tilde{H}}}}^{*}}\mathbf{\tilde{H}\tilde{W}}{{{\mathbf{\tilde{F}}}}^{1\text{D}}}{{\mathbf{Q}}_{l}}} \right)^{-1} \\ 
  & \left ( \lambda {{\mathbf{F}}^{\text{PE,*}}}{{\mathbf{U}}^{*}}\mathbf{Y}_{m}^{\text{PE-P}}+\beta \sum\limits_{n=1}^{N}{\mathbf{R}_{n}^{*}{{{\mathbf{\tilde{H}}}}^{*}}\left( \mathbf{Z}{{_{n}^{\text{PE}}}^{\left( k \right)}}-\frac{\mathbf{D}{{_{n}^{\text{PE}}}^{\left( k \right)}}}{\beta } \right)} + \beta \sum\limits_{l=1}^{L}{\mathbf{Q}_{l}^{*}{{{\mathbf{\tilde{F}}}}^{1\text{D,*}}}{{{\mathbf{\tilde{W}}}}^{*}}{{{\mathbf{\tilde{H}}}}^{*}}\left( \mathbf{Z}{{_{l}^{\text{P}}}^{\left( k \right)}}-\frac{\mathbf{D}{{_{l}^{\text{P}}}^{\left( k \right)}}}{\beta } \right)} \right ).
\end{aligned}
\end{medsize}
\end{equation}

The solution of Eq. \eqref{(SHLR-P_Zpe)} can be obtained by

\begin{equation} \label{(SHLR-P_Zpe_solution)}
\mathbf{Z}{{_{n}^{\text{PE}}}^{\left( k+1 \right)}} = {{S}_{{{\lambda }_{2}}/\beta }}\left( \mathbf{\tilde{H}}{{\mathbf{P}}_{n}}\mathbf{X}{{_{m}^{\text{PE-P}}}^{\left( k+1 \right)}}+\frac{\mathbf{D}{{_{n}^{\text{PE}}}^{\left( k \right)}}}{\beta } \right).
\end{equation}

Similarly, the Eq. \eqref{(SHLR-P_Zt)} can be solved by

\begin{equation} \label{(SHLR-P_Zt_solution)}
\mathbf{Z}{{_{l}^{\text{P}}}^{\left( k+1 \right)}}={{S}_{1/\beta }}\left( \mathbf{\tilde{H}\tilde{W}}{{{\mathbf{\tilde{F}}}}^{1\text{D}}}{{\mathbf{Q}}_{l}}\mathbf{X}{{_{m}^{\text{PE-P}}}^{\left( k+1 \right)}}+\frac{\mathbf{D}{{_{l}^{\text{P}}}^{\left( k \right)}}}{\beta } \right).
\end{equation}

The numerical algorithm for SHLR-VP is summarized in Algorithm \ref{alg:2}.

\begin{algorithm}[htb]
\footnotesize
\caption{Parameter imaging reconstruction with SHLR-VP.}\label{alg:2}
\hspace*{0.02in}{\bf{Input:}} ${{\mathbf{Y}}^{\text{Parameter}}}$, $\mathbf{U}$, $\lambda $, ${{\lambda }_{2}}$, $\beta $, $\tau $.\\
\hspace*{0.02in}{\bf{Output:}} ${\mathbf{X}^\text{Parameter}}$.
\begin{algorithmic}[1]
\WHILE{ $m \le M$ }
\STATE
{\bf{Initialization:}} $\mathbf{Z}_{n}^{\text{PE}}=\mathbf{Z}_{l}^{\text{P}}=\mathbf{0}$, $\mathbf{D}_{n}^{\text{PE}}=\mathbf{D}_{l}^{\text{P}}=\mathbf{1}$, and $k=1$;
\WHILE{ $k\le 100$ and $\left\| \mathbf{X}{{_{m}^{\text{PE-P}}}^{\left( k+1 \right)}}-\mathbf{X}{{_{m}^{\text{PE-P}}}^{\left( k \right)}} \right\|_{F}^{2}/\left\| \mathbf{X}{{_{m}^{\text{PE-P}}}^{\left( k \right)}} \right\|_{F}^{2}\ge {{10}^{-6}}$ }
\STATE
Update $\mathbf{X}{{_{m}^{\text{PE-P}}}^{\left( k \right)}}$ by solving equation \eqref{(SHLR-P_X_solution)};
\STATE
Update $\mathbf{Z}{{_{n}^{\text{PE}}}^{\left( k+1 \right)}}$ and $\mathbf{Z}{{_{l}^{\text{P}}}^{\left( k+1 \right)}}$ by using \eqref{(SHLR-P_Zpe_solution)} and \eqref{(SHLR-P_Zt_solution)};
\STATE
Update multiplier $\mathbf{D}{{_{n}^{\text{PE}}}^{\left( k+1 \right)}}$ and $\mathbf{D}{{_{l}^{\text{P}}}^{\left( k+1 \right)}}$ by using \eqref{(SHLR-P_Dpe)} and \eqref{(SHLR-P_Dt)};
\STATE $ k = k + 1 $;
\ENDWHILE
\STATE $ m = m + 1 $;
\ENDWHILE
\end{algorithmic}
\end{algorithm}

\section{More experimental results for parallel imaging}

In this section, more undersampling patterns and brain images are adopted for further validation in parallel imaging.

We adopted RLNE and MSSIM as evaluation criteria. 
The RLNE is calculated through:
\begin{equation}
\text{RLNE=}\frac{{{\left\| \mathbf{x}-\mathbf{\hat{x}} \right\|}_{2}}}{{{\left\| \mathbf{x} \right\|}_{2}}}.
\end{equation}
where  $\mathbf{x}$ and $\hat{\mathbf{x}}$ denote the vectorized fully sampled and reconstructed SSOS image, respectively. 
The mean measure of structural similarity (MSSIM) on two images $\mathbf{A}$ and $\mathbf{B}$ is calculated by
\begin{equation}
\text{MSSIM} \left( \mathbf{A},\mathbf{B} \right) = \frac{1}{R} \sum\limits_{i=1}^{R}{\frac{\left( 2{{\mu }_{{{\mathbf{a}}_{i}}}}{{\mu }_{{{\mathbf{b}}_{i}}}} + {{C}_{1}} \right) \left( 2{{\sigma }_{{{\mathbf{a}}_{i}}{{\mathbf{b}}_{i}}}} + {{C}_{2}} \right)}{\left( \mu _{{{\mathbf{a}}_{i}}}^{2} + \mu _{{{\mathbf{b}}_{i}}}^{2} + {{C}_{1}} \right) \left( \sigma _{{{\mathbf{a}}_{i}}}^{2} + \sigma _{{{\mathbf{b}}_{i}}}^{2} + {{C}_{2}} \right)}},
\end{equation}%
where $\mathbf{A}$ and $\mathbf{B}$ represent fully sampled and reconstructed SSOS images, respectively; ${{\mu }_{{{\mathbf{a}}_{i}}}}$ , ${{\mu }_{{{\mathbf{b}}_{i}}}}$, ${{\sigma }_{{{\mathbf{a}}_{i}}}}$, ${{\sigma }_{{{\mathbf{b}}_{i}}}}$ and ${{\sigma }_{{{\mathbf{a}}_{i}}{{\mathbf{b}}_{i}}}}$ respectively denote the means, standard deviations and covariance of the local window ${{\mathbf{a}}_{i}}$ and ${{\mathbf{b}}_{i}}$; $R$ the number of local windows. Constants ${{C}_{1}}$ and ${{C}_{2}}$ are introduced to avoid the case when the denominator multiplied by $\mu _{{{\mathbf{a}}_{i}}}^{2}+\mu _{{{\mathbf{b}}_{i}}}^{2}$ and $\sigma _{{{\mathbf{a}}_{i}}}^{2}+\sigma _{{{\mathbf{b}}_{i}}}^{2}$ is close to zero. 

\begin{figure}[htbp]
\setlength{\abovecaptionskip}{0.cm}
\setlength{\belowcaptionskip}{-0.cm}
\centering
\includegraphics[width=6.4in]{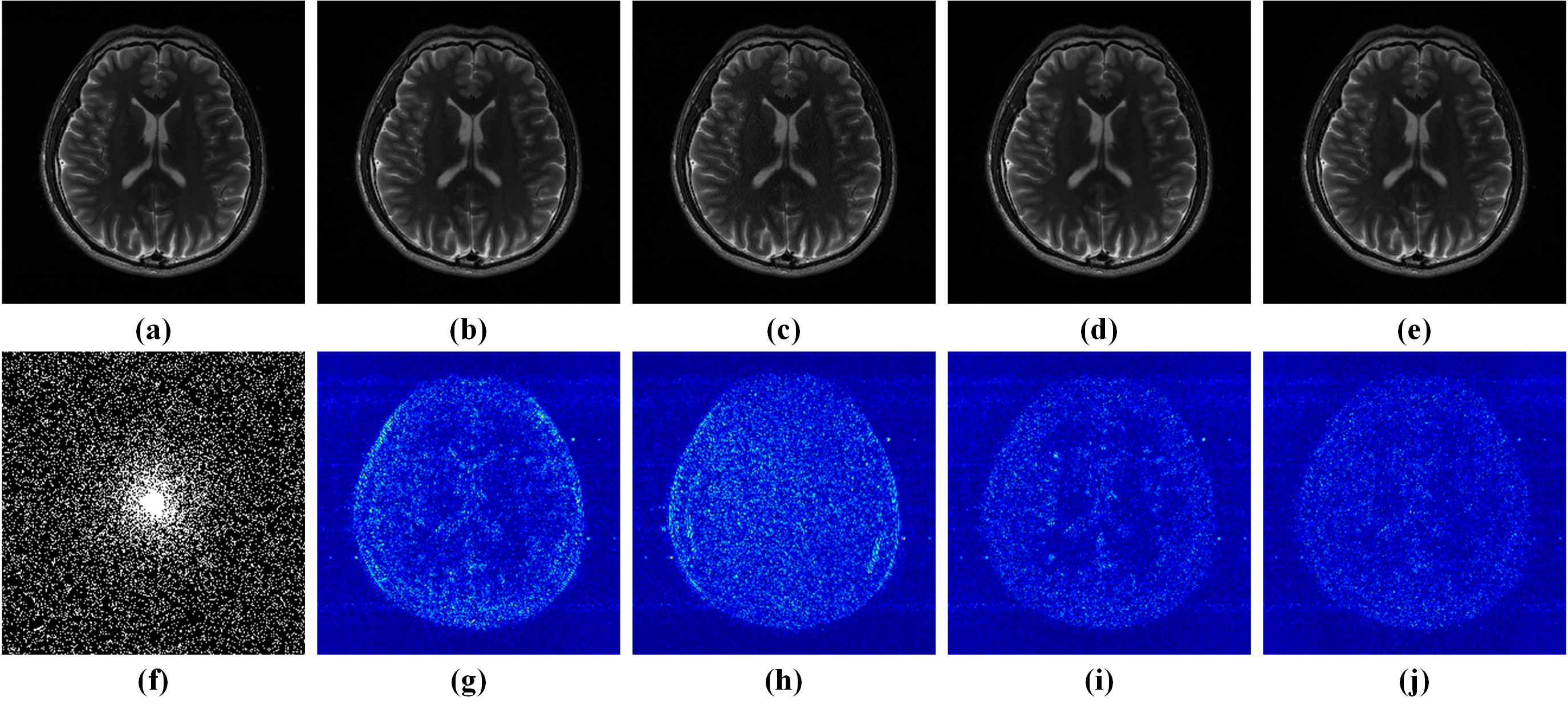}
\caption{Parallel imaging reconstruction results and errors under two patterns. (a) An SSOS image of fully sampled data; (b-e) SSOS images of reconstructed results by $\ell_1$-SPIRiT, AC-LORAKS, STDLR-SPIRiT and SHLR-SV, respectively; (f) the 2D random undersampling pattern of sampling rate $0.18$; (g-j) the reconstruction error distribution ($12.5 \times$) corresponding to reconstructed image above them. Note: the RLNE/MSSIM of (b-e) are $0.0537/0.9878$, $0.0595/0.9767$, $0.0413/ \mathbf{0.9926}$, $\mathbf{0.0396}/0.9925$, respectively.}
\label{fig_SM_2D_random}
\end{figure}

Here, we first validate the proposed method under 2D random undersampling pattern. The 2D random undersampling patterns, which have stronger incoherence, are adopted here to simulate the 3D imaging with two phase encoding. As shown in Figs. \ref{fig_SM_2D_random} (b-e), all these four approaches reconstruct with artifact-free images. The reconstruction image of AC-LORAKS seems relatively more noise, but the error distributes randomly inside the skull, being preferred in clinical application. Errors of $\ell_1$-SPIRiT, STDLR-SPIRiT and SHLR-VC have reached a promising low level, assuring reliable reconstructions. It can also be seen that the SHLR-VC and STLR-SPIRiT provides slightly better reconstruction than that of $\ell_1$-SPIRiT in terms of RLNE and MSSIM.

\begin{figure}[htbp]
\setlength{\abovecaptionskip}{0.cm}
\setlength{\belowcaptionskip}{-0.cm}
\centering
\includegraphics[width=6.4in]{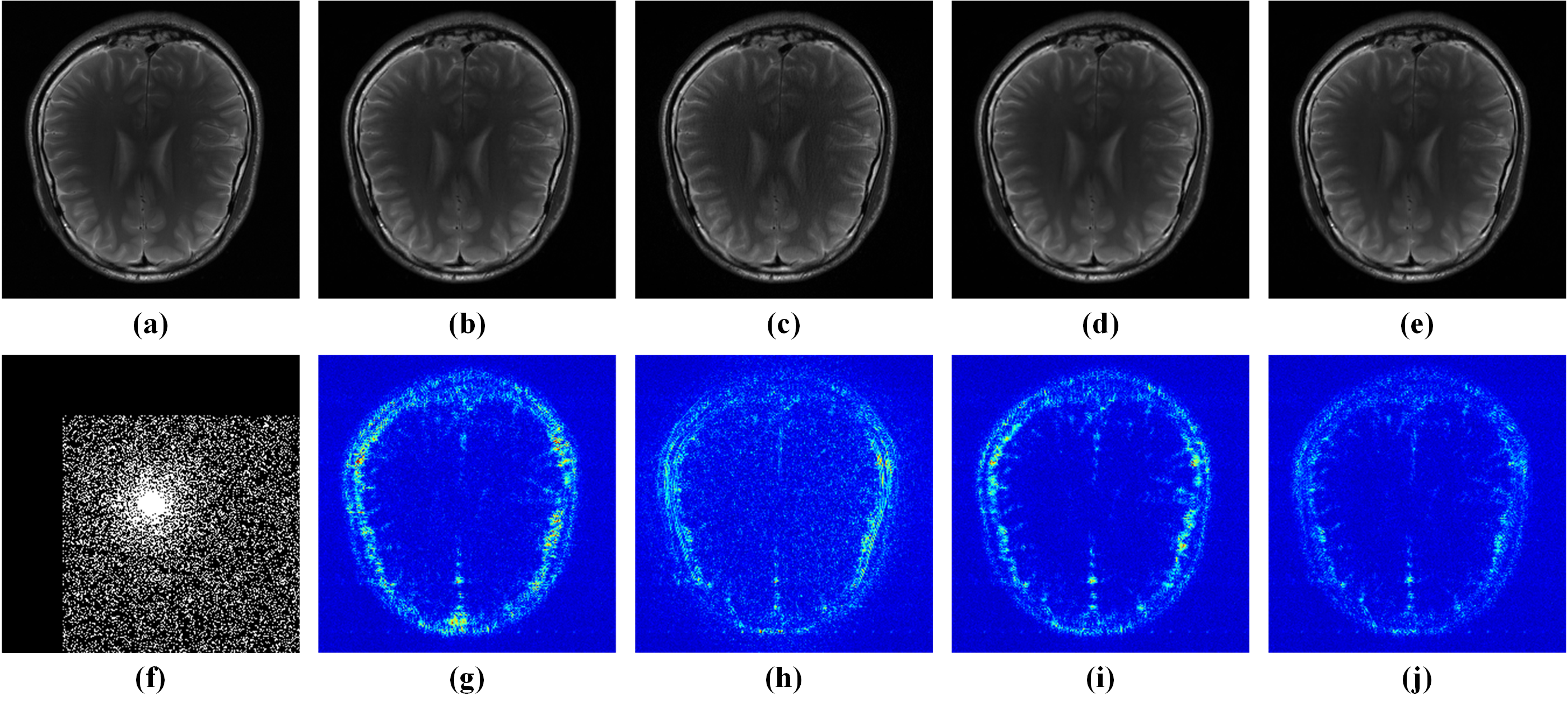}
\caption{Parallel imaging reconstruction results and errors under two patterns. (a) An SSOS image of fully sampled data; (b-e) SSOS images of reconstructed results by $\ell_1$-SPIRiT, AC-LORAKS, STDLR-SPIRiT and SHLR-SV, respectively; (f) the 2D random undersampling pattern with $4/5$ PF (total sampling rate $0.17$); (g-j) the reconstruction error distribution ($10 \times$) corresponding to reconstructed image above them. Note: the RLNE/MSSIM of (b-e) are $0.0493/0.9874$, $0.0489/0.9743$, $0.0420/0.9908$, $\mathbf{0.0362}/ \mathbf{0.9916}$, respectively.}
\label{fig_SM_2D_PF}
\end{figure}

Similarly, SHLR-SV outperforms the comparison methods under 2D random undersampling with PF. As shown in Fig. \ref{fig_SM_2D_PF} (g-j), the proposed method yields the reconstruction with lowest error distribution. The metrics listed in Table \ref{Table_2D_table} also demostrate the superiority of the proposed method.

\begin{figure}[htbp] 
\setlength{\abovecaptionskip}{0.cm}
\setlength{\belowcaptionskip}{-0.cm}
\centering
\includegraphics[width=4.5in]{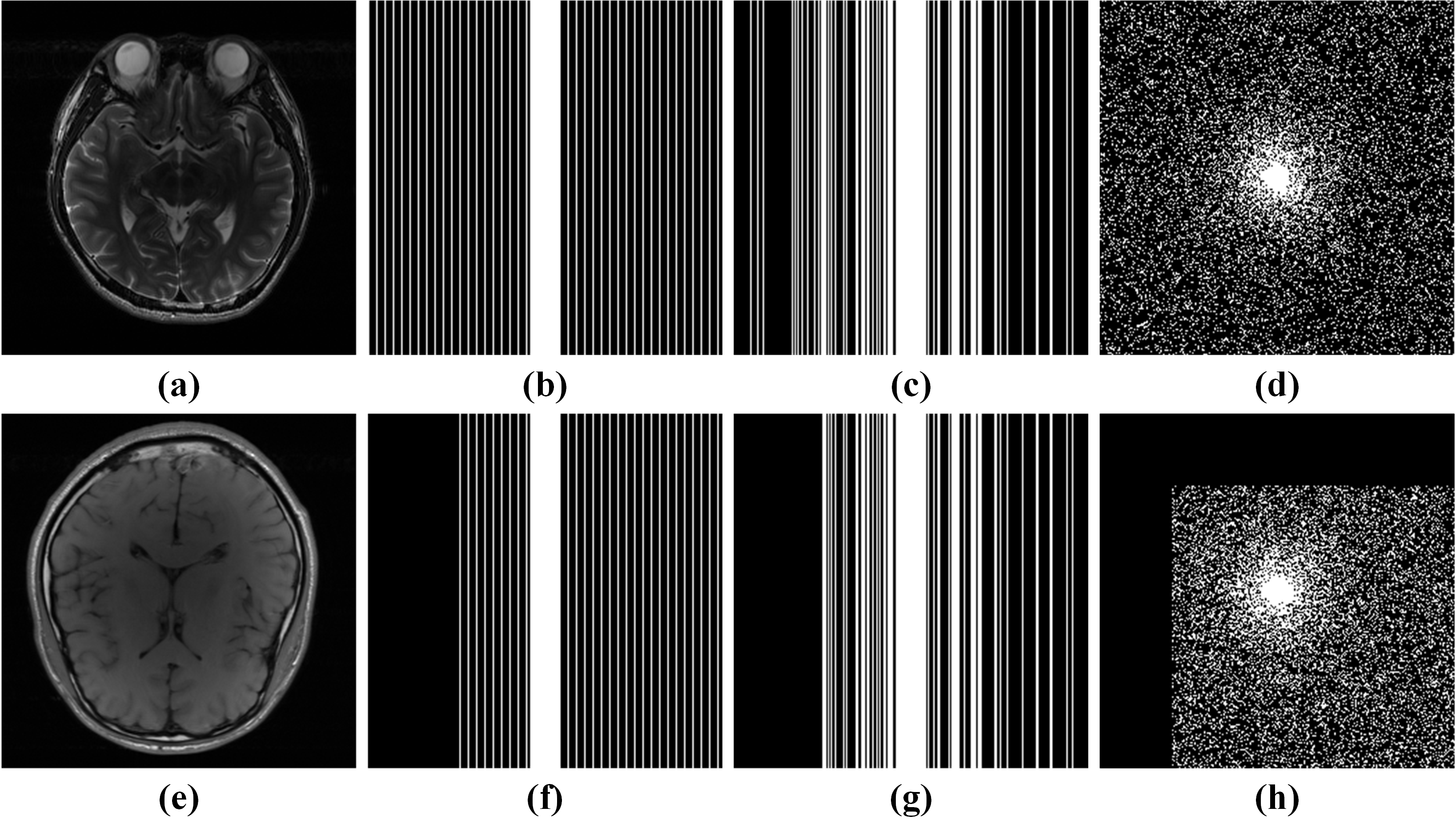}
\caption{More parallel imaging brain images and undersampling patterns. (a) and (e) Two different brain images; (b) the uniform undersampling pattern with $R=6$ and $20$ ACS lines; (c) Cartesian sampling pattern of a sampling rate 0.34; (d) the 2D random un- dersampling pattern of sampling rate 0.18; (e) the $R=6$ uniform undersampling pattern with $20$ ACS lines with $3/4$ PF (total sampling rate $0.20$); (f) the Cartesian undersampling pattern with $3/4$ PF (total sampling rate $0.30$); (j) the 2D random undersampling pattern with $4/5$ PF (total sampling rate $0.17$.)}
\label{fig_2D_dataset}
\end{figure}

Besides, we adopted two more \textit{in vivo} brain data under six different undersampling patterns to demonstrate the performance of the proposed method (shown in Fig. \ref{fig_2D_dataset}). Two brain datasets acquired from healthy volunteer. The first dataset shown in Fig. \ref{fig_2D_dataset} (a) is acquired from a 3T SIEMENS MRI scanner (Siemens Healthcare, Erlangen, Germany) equipped with 32 coils using T2-weighted turbo spin echo sequence (matrix size = $256 \times 256$, TR/TE = $6100 / 99$ ms, FOV = $220$ mm $\times$ $220$ mm, slice thickness = $3$ mm). Four virtual coils are compressed from the acquired data of 32 coils \cite{2013_Coil_Compression}. The other dataset shown in Fig. \ref{fig_2D_dataset} (e) is obtained from a 3T SIEMENS MRI scanner (Siemens Healthcare, Erlangen, Germany) equipped with  32 coils using the 2D T1-weighted FLAIR sequence (matrix size = $256 \times 256$, TR/TE = $3900 / 9.3$ ms, FOV = $200$ mm $\times$ $200$ mm, slice thickness = $5$ mm). Eight virtual coils are compressed from the acquired data of 32 coils \cite{2013_Coil_Compression}.

The quality metric of reconstruction results are listed in Table \ref{Table_2D_table}. The same phenomenon as that in the main text can be observed. The proposed method in parallel imaging, SHLR-SV, is superior to $\ell_1$-SPIRiT and AC-LORAKS, and comparable to the STDLR-SPIRiT, in terms of RLNE and MSSIM. When it turns to undersampling patterns with PF, SHLR-SV performs slightly better results than STDLR-SPIRiT does.

\begin{table}[htbp]
  \centering
  \caption{RLNE/MSSIM results for brains in Fig. \ref{fig_2D_dataset} using different undersampling patterns.}
    \begin{tabular}{m{1.5cm}<{\centering}m{1.5cm}<{\centering}m{2.5cm}<{\centering}m{2.5cm}<{\centering}m{2.5cm}<{\centering}m{2.5cm}<{\centering}}
    \toprule
    Image & Pattern & $\ell_1$-SPIRiT & AC-LORAKS & STDLR-SPIRiT & SHLR-SV \\
    \midrule
    \multicolumn{1}{c}{\multirow{6}[2]{*}{Fig. \ref{fig_2D_dataset} (a)}} & Fig. \ref{fig_2D_dataset} (b) & 0.1062/0.9608 & 0.0710/0.9747 & 0.0713/0.9794 & \textbf{0.0611}/\textbf{0.9863} \\
          & Fig. \ref{fig_2D_dataset} (c) & 0.0810/0.9794 & 0.0675/0.9793 & 0.0599/0.9883 & \textbf{0.0531}/\textbf{0.9901} \\
          & Fig. \ref{fig_2D_dataset} (d) & 0.0704/0.9800 & 0.0698/0.9732 & 0.0551/0.9871 & \textbf{0.0530}/\textbf{0.9870} \\
          & Fig. \ref{fig_2D_dataset} (f) & 0.1300/0.9508 & 0.0858/0.9655 & 0.0977/0.9721 & \textbf{0.0699}/\textbf{0.9832} \\
          & Fig. \ref{fig_2D_dataset} (g) & 0.1101/0.9688 & 0.0920/0.9629 & 0.0884/0.9809 & \textbf{0.0698}/\textbf{0.9849} \\
          & Fig. \ref{fig_2D_dataset} (h) & 0.0906/0.9870 & 0.0729/0.9730 & 0.0797/0.9840 & \textbf{0.0570}/\textbf{0.9871} \\
    \midrule
    \multicolumn{1}{c}{\multirow{6}[2]{*}{Fig. \ref{fig_2D_dataset} (e)}} & Fig. \ref{fig_2D_dataset} (b) & 0.0722/0.9578 & 0.0613/0.9446 & \textbf{0.0425}/0.9811 & 0.0499/\textbf{0.9820} \\
          & Fig. \ref{fig_2D_dataset} (c) & 0.0397/0.9861 & 0.0410/0.9753 & 0.0298/0.9915 & \textbf{0.0297}/\textbf{0.9918} \\
          & Fig. \ref{fig_2D_dataset} (d) & 0.0292/0.9897 & 0.0412/0.9619 & \textbf{0.0246}/0.9920 & 0.0254/\textbf{0.9923} \\
          & Fig. \ref{fig_2D_dataset} (f) & 0.0804/0.9505 & 0.0652/0.9402 & 0.0493/0.9797 & \textbf{0.0483}/\textbf{0.9809} \\
          & Fig. \ref{fig_2D_dataset} (g) & 0.0506/0.9823 & 0.0503/0.9698 & 0.0387/0.9895 & \textbf{0.0365}/\textbf{0.9897} \\
          & Fig. \ref{fig_2D_dataset} (h) & 0.0419/0.9870 & 0.0401/0.9777 & 0.0363/0.9911 & \textbf{0.0308}/\textbf{0.9920} \\
    \bottomrule
    \end{tabular}%
  \label{Table_2D_table}%
\end{table}%

\newpage

\section{More experimental results for parameter imaging}
In this section, we adopted two \textit{in vivo} T2 mapping data to verify the performance of the proposed method.

We first test the comparison methods under the sampling pattern with partial Fourier. The dataset used here was fully sampled on a 3T SIEMENS MRI scanner (Siemens Healthcare, Erlangen, Germany) with 20 coil using turbo spin echo sequence (TR = $4000$ ms, $15$ TE with $8.2$, $16$, $25$, $33$, $41$, $49$, $58$, $66$, $74$, $82$, $91$, $99$, $107$, $115$, and $124$ ms, FOV = $220$ mm $\times$ $220$ mm, matrix size = $192 \times 192$, slice thickness = $4$ mm). We removed 4 coils of data with strong noise, and finally, 16 coils of data were used for reconstructions.

\begin{figure*}[htbp]
\setlength{\abovecaptionskip}{0.cm}
\setlength{\belowcaptionskip}{-0.cm}
\centering
\includegraphics[width=6.0in]{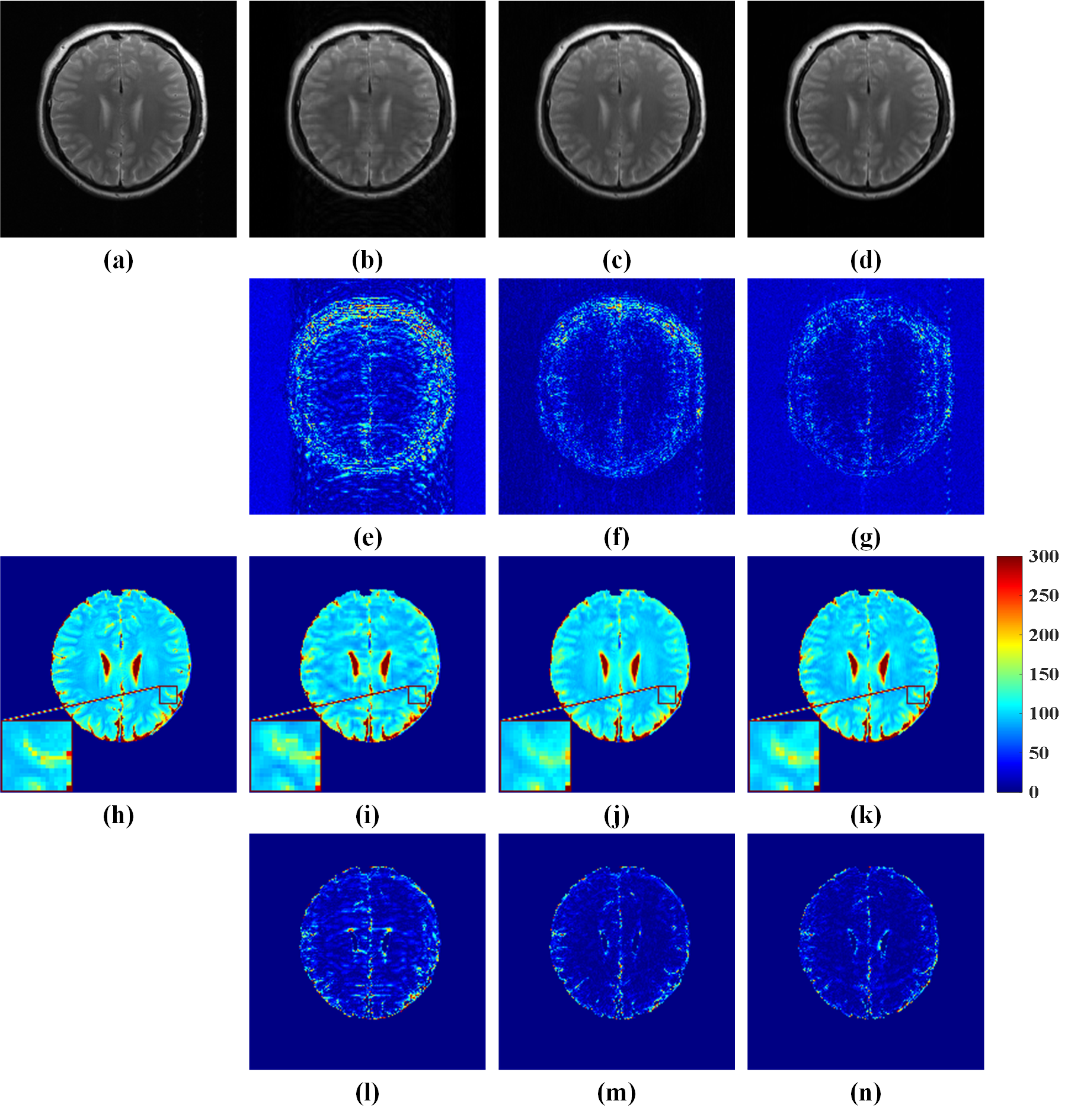}
\caption{T2 mapping reconstruction results and errors under $R=8$ undersampling pattern with PF. (a) The ($9$-th echo) images of fully sampled data; (b-d) reconstructed images of MORASA, ALOHA, and SHLR-VP, respectively; (e-g)) the reconstruction error distribution ($8 \times$) corresponding to reconstructed image above them; (h) T2 map of fully sampled data; (i-k) T2 maps of reconstructed results at R=8 by MORASA, ALOHA, and SHLR-VP, respectively; (l-n) the reconstruction error distribution ($10 \times$) corresponding to reconstructed image above them. Note: the average RLNE/MSSIM of reconstructed images of MORASA, ALOHA, and SHLR-VP are $0.0967/0.9577$, $0.0599/0.9807$, and $\mathbf{0.0563} / \mathbf{0.9828}$, respectively. The RLNE/MSSIM of reconstructed T2 maps of MORASA, ALOHA, and SHLR-VP are $0.1404/0.9794$, $0.0983/0.9889$, and $\mathbf{0.0966} / \mathbf{0.9904}$, respectively.}
\label{fig_mapping_PF}
\end{figure*}

Look at the reconstructed image at $9$-th echo, we can find that the MORASA image has obvious sampling artifacts (Fig. \ref{fig_mapping_PF} (b) and (e)). The artifacts can still be seen in its mapping results (Fig. \ref{fig_mapping_PF} (i) and (I)). Both ALOHA and SHLR-VP remove the sampling artifacts promisingly (Fig. \ref{fig_mapping_PF} (c) and (d)). However, we can still see significant difference between their error images (Fig. \ref{fig_mapping_PF} (f) and (g)). The SHLR-VP error is much smaller than ALOHA. The associated mapping result of SHLR is more consistent with that of ALOHA (Please see the zoom-in image in Fig. \ref{fig_mapping_PF} (j) and (k)).

Besides, we conducted experiments on undersampling patterns with different acceleration factors and patterns with/without PF. Another slice of T2 mapping data shown in Fig. 10 in the main text was adopted here. The results are summerized in Table \ref{Table_mapping_table}. As shown in Table \ref{Table_mapping_table}, the proposed method for parameter imaging, SHLR-VP, yields reconstruction with lower error and better structural preservation than that of comparison methods in both reconstructed images and estimated T2 maps. Especially when the acceleration factor is higher ($R=8$) or when the undersampling patterns with PF, the improvement gained by SHLR-VP appears more obvious.

\begin{table}[htbp]
  \centering
  \caption{RLNE/MSSIM of parameter imaging reconstruction under different undersampling patterns.}
    \begin{tabular}{m{3cm}<{\centering}m{3cm}<{\centering}m{3cm}<{\centering}m{3cm}<{\centering}}
    \toprule
    Pattern & MORASA & ALOHA & SHLR-VP \\
    \midrule
    \multicolumn{4}{p{12cm}}{Average RLNEs of reconstructed images} \\
    $R=6$    & 0.0730/0.9844 & 0.0621/0.9819 & \textbf{0.0611}/\textbf{0.9882} \\
    $R=8$    & 0.0840/0.9812 & 0.0793/0.9775 & \textbf{0.0718}/\textbf{0.9843} \\
    $R=6$ with PF & 0.0840/0.9793 & 0.0776/0.9752 & \textbf{0.0722}/\textbf{0.9836} \\
    $R=8$ with PF & 0.0967/0.9718 & 0.0929/0.9698 & \textbf{0.0815}/\textbf{0.9802} \\
    \midrule
    \multicolumn{4}{p{12cm}}{RLNEs of reconstructed T2 maps} \\
    $R=6$    & 0.1080/0.9881 & \textbf{0.0917}/0.9897 & 0.0921/\textbf{0.9909} \\
    $R=8$    & 0.1182/0.9852 & 0.1201/0.9836 & \textbf{0.1077}/\textbf{0.9873} \\
    $R=6$ with PF & 0.1309/0.9830 & 0.1370/0.9790 & \textbf{0.1113}/\textbf{0.9861} \\
    $R=8$ with PF & 0.1423/0.9783 & 0.1458/0.9762 & \textbf{0.1147}/\textbf{0.9854} \\
    \bottomrule
    \end{tabular}%
  \label{Table_mapping_table}%
\end{table}%

\newpage

\section{Parameter selection of the proposed approaches}
In this subsection, the effect of parameters setting of the two proposed methods will be discussed, including regularization parameters ${{\lambda }}$ and ${{\lambda }_{1}}$ for parallel imaging and ${{\lambda }}$ and ${{\lambda }_{2}}$ for parameter imaging.
The brain image in Fig. 8 (a) and the sampling pattern in Fig. 8 (f) has been used to perform parallel imaging reconstructions, and the T2 mapping data shown in Fig. 10 with the $R=8$ reduction factor has been adopted for parameter imaging reconstruction.

\begin{figure}[htbp]
\setlength{\abovecaptionskip}{0.cm}
\setlength{\belowcaptionskip}{-0.cm}
\centering
\includegraphics[width=6.0in]{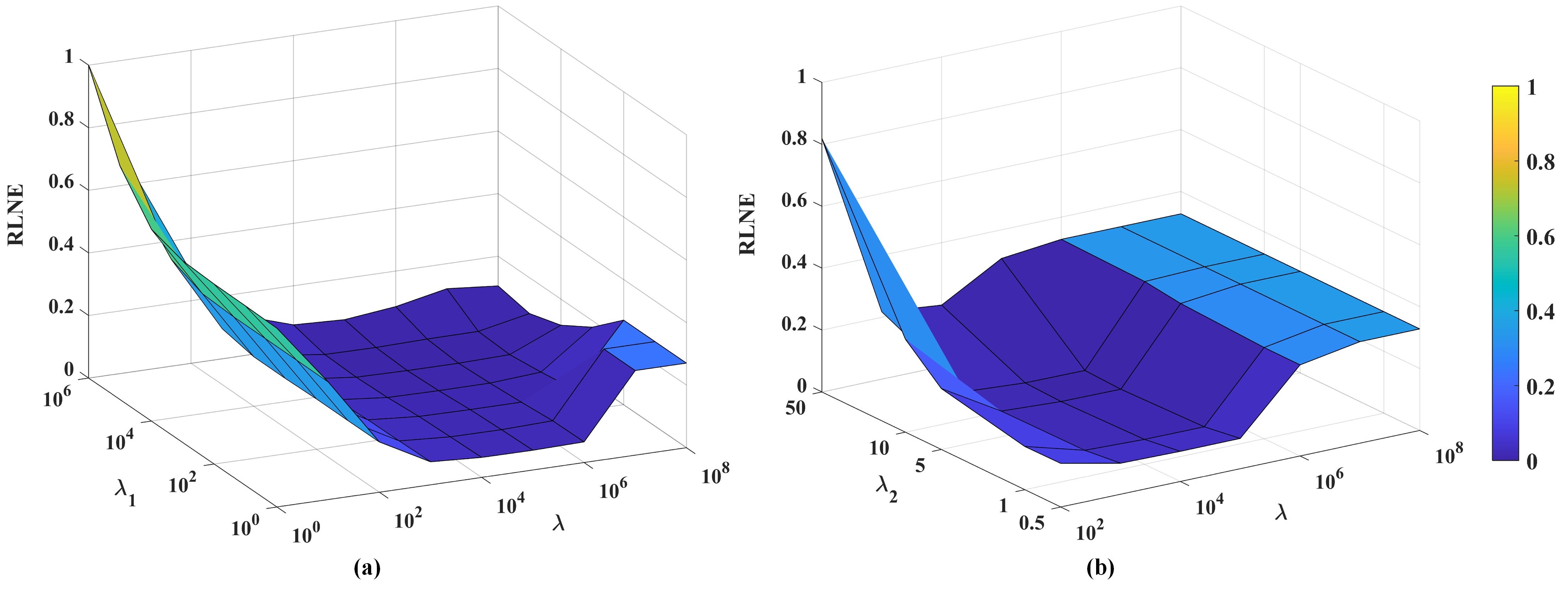}
\caption{RLNEs versus regularization parameters of SHLR-SV and SHLR-VP. (a) RLNEs versus $\lambda$ and ${\lambda}_{1}$ in SHLR-SV, (b) RLNEs versus ${\lambda}$ and ${\lambda }_{2}$ in SHLR-VP.}
\label{fig_lambda}
\end{figure}

The reconstruction errors versus different values of regularization parameters of SHLR-SV are shown in Fig. \ref{fig_lambda} (a). As can be seen in Fig. \ref{fig_lambda} (a), there exists a wide range of ${{\lambda_{1}}}$ (${{10}^{0}}\le {{\lambda}_{1}}\le {{10}^{4}}$) and ${{\lambda}}$ (${{10}^{3}}\le {{\lambda}}\le {{10}^{6}}$) leading to relatively low reconstruction errors. Too small or too large values of ${{\lambda}}$ and ${{\lambda}_{2}}$ produce higher reconstruction errors. Similar trend was observed in the parameter imaging results (Fig. \ref{fig_lambda} (b)) but the ranges of $\lambda$ and $\lambda_{2}$ are smaller. There exists a range of ${{\lambda_{2}}}$ ($1 \le {{\lambda}_{2}}\le 5$) and ${{\lambda}}$ (${{10}^{3}}\le {{\lambda}}\le {{10}^{5}}$) leading to relatively low reconstruction errors.

%----------------------------------------------------------------------
%----------------------------- References -----------------------------
%----------------------------------------------------------------------
\ifCLASSOPTIONcaptionsoff
  \newpage
\fi

\bibliographystyle{IEEEtran}
\bibliography{IEEEabrv,Mylib}